\pdfoutput=1

\documentclass[nonacm]{acmart}
\AtBeginDocument{%
  }

\usepackage{float}
\usepackage{enumitem}
\usepackage[official]{eurosym}
\usepackage{colortbl} 
\usepackage{multirow}
\usepackage{xspace}
\usepackage{longtable}

\newcommand{\amsalgorithm}{\textit{AMS algorithm}\xspace}

\definecolor{GroupGreen}{HTML}{88c946}
\definecolor{GroupBlue}{HTML}{50cfe2}
\definecolor{GroupRed}{HTML}{FF8E72}
\definecolor{GroupYellow}{HTML}{ffdb57}
\definecolor{OliveGreen}{HTML}{3f7e31}
\definecolor{BrickRed}{HTML}{c43119}

\renewcommand\footnotetextcopyrightpermission[1]{} % removes footnote with conference information in first column
\setcopyright{none}
\copyrightyear{2025}
\acmYear{2025}
\acmDOI{XXXXXXX.XXXXXXX}

\begin{document}

% Old title: Deliberative XAI: How Explanations Impact Understanding and Decision-Making of AI Novices
% New title: Understanding and Deliberating: How explanations help groups and individuals learn and form opinions about AI systems
% Designing Explanations to Support AI Novices in Understanding and Making Decisions about AI Systems Alone and in Groups
% Better Together? The Role of Explanations in Supporting Novices in Individual and Collective Deliberations about AI (Laura)
\title[The Role of Explanations in Supporting Novices in Individual and Collective Deliberations about AI]{Better Together? The Role of Explanations in Supporting Novices in Individual and Collective Deliberations about AI}

% Remove contact info from first page
\makeatletter
\let\@authorsaddresses\@empty
\makeatother

\author{Timothée Schmude}
\email{timothee.schmude@univie.ac.at}
\orcid{0009-0006-8276-4670}
\affiliation{%
  \institution{University of Vienna, Faculty of Computer Science, Research Network Data Science, Doctoral School Computer Science}
  \city{Vienna}
  \country{Austria}
}

\author{Laura Koesten}
\email{laura.koesten@univie.ac.at}
\orcid{XXXX}
\affiliation{%
  \institution{Mohamed bin Zayed University of Artificial Intelligence, Department of Human-Computer Interaction}
  \city{Abu Dhabi}
  \country{UAE}
}
\affiliation{%
  \institution{and University of Vienna, Faculty of Computer Science, Research Group Visualization and Data Analysis}
  \city{Vienna}
  \country{Austria}
}
\affiliation{%
  \institution{AIT Austrian Institute of Technology GmbH, Center for Technology Experience}
  \city{Vienna}
  \country{Austria}
}

\author{Torsten Möller}
\email{torsten.moeller@univie.ac.at}
\orcid{XXXX}
\affiliation{%
  \institution{University of Vienna, Faculty of Computer Science, Research Network Data Science, Research Group Visualization and Data Analysis}
  \city{Vienna}
  \country{Austria}
}

\author{Sebastian Tschiatschek}
\email{sebastian.tschiatschek@univie.ac.at}
\orcid{XXXXX}
\affiliation{%
  \institution{University of Vienna, Faculty of Computer Science, Research Network Data Science, Research Group Data Mining and Machine Learning}
  \city{Vienna}
  \country{Austria}
}

\renewcommand{\shortauthors}{Schmude et al.}

% What is the specific problem?
% What have you done?
% What are the implications on the bigger picture?
% or
% Importance
% Problem
% Aims and approach (we [aim] using [method])
% Key findings
% Conclusion / significance
% 150-200 word
% TOCHI: short, direct, and complete sentences. It should be informative enough to serve in some cases as a substitute for reading the paper itself. It should state the objectives of the work, summarize the results, and give the principle conclusions,
\begin{abstract}

  Deploying AI systems in public institutions can have far-reaching consequences for many people, making it a matter of public interest. Providing opportunities for stakeholders to come together, understand these systems, and debate their merits and harms is thus essential. Explainable AI often focuses on individuals, but deliberation benefits from group settings, which are underexplored. To address this gap, we present findings from an interview study with 8 focus groups and 12 individuals. Our findings provide insight into how explanations support AI novices in deliberating alone and in groups. Participants used modular explanations with four information categories to solve tasks and decide about an AI system’s deployment. We found that the explanations supported groups in creating shared understanding and in finding arguments for and against the system's deployment. In comparison, individual participants engaged with explanations in more depth and performed better in the study tasks, but missed an exchange with others.  Based on our findings, we provide suggestions on how explanations should be designed to work in group settings and describe their potential use in real-world contexts. With this, our contributions inform XAI research that aims to enable AI novices to understand and deliberate AI systems in the public sector.

\end{abstract}

\begin{CCSXML}
<ccs2012>
   <concept>
       <concept_id>10003120.10003121.10003122.10003334</concept_id>
       <concept_desc>Human-centered computing~User studies</concept_desc>
       <concept_significance>500</concept_significance>
       </concept>
 </ccs2012>
\end{CCSXML}

\ccsdesc[500]{Human-centered computing~User studies}

\keywords{explainable AI, understanding, deliberation, qualitative methods, focus groups}


%% Published manuscripts are typically 20-60 pages — with rare exceptions on either side of that range.
\maketitle

\section{Introduction}
\label{sec:introduction}

% 1. What is the problem?

A growing number of AI systems\footnote{We use the term `AI system' to describe algorithmic systems with machine learning components. The nomenclature follows research on explainable AI~\cite{langer_what_2021} and research on AI in the context of society~\cite{zuger_ai_2023, collins_right_2024} and regulation~\cite{panigutti2023}.} are deployed in the public sector to decide about critical issues, such as employment, migration, and criminal justice~\cite{scott_algorithmic_2022, Bansak2018, zuger_ai_2023, chouldechova2017}. These systems can have consequences for all stakeholders but tend to have the largest impact on their decision subjects (people the system decides over), such as discrimination or misclassification~\cite{brown_toward_2019, raji_fallacy_2022}. These harms intensify when decision-making is opaque and uncontestable~\cite{ananny_seeing_2018, alfrink2023, definelicht2020}. For these reasons, public AI systems should be considered as `matters of public interest'~\cite{zuger_ai_2023}, meaning that they need to be explainable, justifiable, and open for public deliberation~\cite{kawakami2024, soton2024, brown_toward_2019}. Explanations can make AI systems more understandable and easier to assess and control~\cite{langer_what_2021}. Much of explainable AI (XAI) research is focused on individuals, but research has shown that group settings can facilitate the understanding of complex topics~\cite{moshman_collaborative_1998, nokes-malach_when_2015, navajas_aggregated_2018}. Further, group settings encourage the exchange of views and arguments~\cite{stromer-galley_measuring_2007, smith_why_2009}, which are vital when engaging in deliberation (collectively finding a solution to a problem~\cite{habermas1991structural}). XAI has not explored in detail how explanations can be combined with group settings to leverage these benefits. In this paper, we aim to address this gap.

% 2. Why is it interesting and important?
As arguably many people affected by decisions of AI systems have no technical background in developing or analyzing these systems, this work focuses on `lay people'~\cite{lima2023, devito2018, shulner-tal_enhancing_2022} or `\textit{AI novices}'~\cite{mohseni2020}. Explanations for AI novices naturally have different requirements than explanations for AI practitioners, as they have different expertise~\cite{ehsan_who_2024}, interests~\cite{langer_what_2021}, and prior knowledge~\cite{szymanski_visual_2021, cheng_explaining_2019, schmude2024information}. Explanation formats (e.g., visual, textual, dialogue) are known to impact AI novices' understanding~\cite{cheng_explaining_2019, bove2022, szymanski_visual_2021} but show inconsistent effects~\cite{bove2022} due to contextual factors such as participants' perceptions of the use case domain. A possible solution is the adaptation of explanations with personalization~\cite{shulner-tal_enhancing_2022, conati_toward_2021, naiseh2020_personalising} and interactivity~\cite{cheng_explaining_2019, bertrand2023, guesmi_interactive_2023}. To this end, previous research has analyzed AI novices' information needs~\cite{schmude2024information} and developed explanations in collaboration with end-users~\cite{weitz2024, lee_webuildai_2019}. However, designs that can adapt to AI novices and that support their deliberation, especially in groups, are still rare~\cite{chiang_enhancing_2024, soton2024}. But these explanations are essential to provide opportunities where AI novices can learn about and discuss AI systems and to realize the principles of human-centered AI (engaging stakeholders and empowering people~\cite{ShneidermanBen2022HA}).   
%Psychological research has reported mixed results on whether groups perform better in tasks than individuals, finding both ``process gain'' and ``process loss'' from group interactions~\cite{kerr_group_2004}. 

% 3. Why is it hard? (E.g. why do naive approaches fail?)
Designing explanations that support understanding and deliberation for AI novices in both group and individual settings meets multiple challenges. Group composition and dynamics place special demands on explanation design~\cite{naiseh_explainable_2021}, as explanations need to cater to a diverse set of information and format preferences~\cite{schmude2024information, bove_why_2024, bertrand2023}. They must further support a joint understanding process and collaborative interactions~\cite{long_role_2021}, such as sharing and combining, all while providing comprehensive information and remaining clear and navigable. We address these challenges by proposing a modular explanation design that spans four information categories (\textit{data, system details, usage, and context}) from which users can select. Another challenge consists in validating explanation approaches qualitatively with the relevant stakeholder groups. Specifically, XAI research does not always include people from marginalized population groups, who are most likely to be affected negatively as decision subjects~\cite{brown_toward_2019}. To address this, we conducted two focus groups with decision subjects to include their perspectives and voices on AI systems in the public sector.      

% 4. Why hasn't it been solved before? (Or, what's wrong with previous proposed solutions? How does mine differ?)
To examine the role of explanations in supporting AI novices' understanding and deliberation we present the findings of a task-based interview study with 43 participants, involving 8 focus groups and 12 single interviews. For this study, we used an explanation design comprising 36 single explanations in question-answer pairs. These explanations are organized into the four categories \textit{data}, \textit{system details}, \textit{usage}, and \textit{context} and further assigned to subtopics and levels of detail (\autoref{fig:explanation_overview}). Participants used these explanations to solve the study tasks and decide about deploying a public AI system (\autoref{fig:study_procedure}). We used an employment scoring algorithm that connects to previous work on AI systems in employment~\cite{scott_algorithmic_2022, niklas2015, lopez_reinforcing_2019}. Our analysis examines participants' self-reported understanding, decision confidence, and perceptions of key information. We further conducted a thematic analysis of how participants interacted with explanations in both settings. The following research questions guide the analysis:

\begin{enumerate}[label={[RQ\arabic*]}, leftmargin=10ex]
    \item \emph{Explanations}: How does a question-driven, modular explanation design support AI novices’ understanding in groups and individual settings? 
    \item \emph{Deliberation}: How do AI novices use explanations to form opinions and make decisions about AI systems?
\end{enumerate}

% 5. What are the key components of my approach and results? Also include any specific limitations. (Contributions!)
Our contributions include i) an explanation design that builds on a question-driven and modular design to accommodate different levels of completeness and soundness and that is suitable for both individual and group settings; ii) an in-depth description of how explanations support participants' understanding and deliberation processes that identifies salient differences between the two settings; iii) an analysis of which type of explanations participants requested most often and perceived as most important; and iv) recommendations regarding the design and use of explanations in group settings. We envision that this work can provide valuable starting points for future XAI research that aims to connect explanations to deliberation on public AI systems.

\section{Background and Related Work}

This section describes how our work is embedded in human-centered explainable AI and outlines the main challenges and approaches to designing explanations for AI novices. It further introduces the two main lenses of analysis to answer our research questions: understanding and deliberation.   

\subsection{Human-centered explainable AI}

% What is it again and why is our work situated in it and why should we care
% Establish XAI, give motivation for why we focus on AI novices so much

Explainability is often described as a cornerstone of responsible AI systems~\cite{thiebes_trustworthy_2021}, as explanations can enable stakeholders such as users and decision subjects to understand~\cite{langer_what_2021} and contest AI decisions~\cite{alfrink2023}. A similar focus is set by the domain of human-centered AI~\cite{capel_what_2023}, which proposes to build AI systems that 1) are based on user-experience design and stakeholder engagement, and 2) empower rather than replace people by being controllable and autonomy-preserving~\cite{ShneidermanBen2022HA, shin_algorithms_2023, xu2019}.
These principles become especially important in high-risk settings~\cite{european_commission_laying_2021}, such as employment~\cite{scott_algorithmic_2022, flugge_perspectives_2021}, immigration~\cite{Bansak2018}, or criminal justice~\cite{chouldechova2017}, where erroneous or non-transparent algorithmic decisions can cause severe harm to those affected~\cite{raji_fallacy_2022}. In response to these risks, the domain of \textit{human-centered explainable AI} (HCXAI) examines how explanations can contribute to ``equitable and ethical Human-AI interaction''~\cite{ehsan_human-centered_2023}. It assumes that transparency alone is not enough to make AI systems explainable~\cite{ananny_seeing_2018}, but that explanations need to consider the system's social context~\cite{wenzelburger_algorithms_2022}, its lifecycle~\cite{dhanorkar_who_2021} and its different stakeholder groups~\cite{ehsan_human-centered_2023}. In the context of this work, human-centered explainability is realized by testing and validating a design approach intended to support AI novices in understanding AI systems and deciding about their deployment in public institutions~\cite{zuger_ai_2023}. 
%With this approach, our work aims to examine suitable formats for explanations that can accompany the responsible deployment of AI systems in public institutions~\cite{zuger_ai_2023} and high-risk settings~\cite{golpayegani_be_2023}. 

\subsection{Designing explanations for AI novices}
\label{sec:related_work_ai_novices}

% Outline who AI novices are, give an overview of what has been done before, describe how we tackle the challenges that others have not
% realized by testing and validating a design approach that has been developed based on qualitative exploration of AI novices’ information needs [95].

The majority of people who interact and are involved with AI systems in public institutions are lay people or \textit{AI novices}, here defined as ``users who [might] use AI products in daily life but have no (or very little) expertise on machine learning systems''~\cite{mohseni2020}. Established explanation methods, like LIME~\cite{Ribeiro2016}, SHAP~\cite{lundberg2017}, and surrogate models~\cite{molnar2022} are tailored to experts and require technical knowledge; hence, they do not address the needs of AI novices. To better cover these needs and match information to them, it is necessary to understand how non-experts conceive of AI systems. Previous HCI research has analyzed lay understandings to explore user perception and understanding of several algorithmic systems~\cite{devito2018, eslami2016}. Similarly, XAI research has begun to explore the information needs of AI novices to design suitable explanations for a broader audience~\cite{schmude2024information}. However, few explanation designs have been proposed that truly assume the perspectives of AI novices~\cite{szymanski_visual_2021, cheng_explaining_2019}. In the following, we summarize current approaches with respect to AI novices' information needs and current practices of explanation design. 

Regarding \textbf{information needs}, previous qualitative research outlined that AI novices value information about the context and intention of a system's deployment~\cite{schmude2024information, kawakami2024} as well as about the responsible institution~\cite{brown_toward_2019}. In contrast to traditional XAI approaches, which focus on descriptive information about the system's workings and outputs, explanations for AI novices thus also require normative information, such as justifications~\cite{biran2017_human_centric_justifications} for design choices. Regarding \textbf{information coverage}, previous work posits that transparency does not equal understanding~\cite{ananny_seeing_2018} and that simply making all information about a system available is no valid explanation strategy. Empirical evaluations of this claim showed that ``white-box'' explanations (transparent models) can improve ``objective''\footnote{We use the term in reference to~\citet{cheng_explaining_2019} and~\citet{bove2022}, it means to describe factual or testable understanding.} understanding but may overwhelm non-expert users and reduce perceived understanding~\cite{cheng_explaining_2019}. However, later work~\cite{bove_why_2024} used similar explanations and found that they had the opposite effect on understanding, attributed to a difference in the studies' use case domains (student admission vs. finance). This indicates that the amount of information should be adaptable to the given context. Regarding \textbf{explanation format}, \citet{szymanski_visual_2021} examined how expert and lay users rated explanations of different formats and found that while lay users favored visual explanations, they performed better with textual ones. Other studies confirm this discrepancy and posit that comprehension varies with demographic factors and domain knowledge~\cite{wang21, shulner-tal_enhancing_2022, ehsan_who_2024}. These issues are assumed to be addressed with \textbf{personalization of explanations}~\cite{conati_toward_2021, shulner-tal_enhancing_2022}, meaning that they are selected and designed according to the user's stakeholder role~\cite{langer_what_2021}, prior knowledge~\cite{schmude2024information}, beliefs~\cite{miller_explanation_2019}, and explanatory stance~\cite{byrne_good_2023, keil2006}. Further aspects to be considered include the explanation's purpose~\cite{freiesleben2023} and the user's familiarity with AI~\cite{kramer2018}. 
These approaches guided the conceptual development of the explanation design presented in this study.
%that addresses the information needs of AI novices~\cite{schmude2024information}, supporting their understanding~\cite{langer_what_2021} and decision-making~\cite{bertrand2023} in individual and collective settings. 

We compiled information from different sources documenting employment prediction algorithms~\cite{scott_algorithmic_2022, allhutter_bericht_ams-algorithmus_2020}, producing an extensive collection of ``scavenged''~\cite{wieringa_hey_2023} material. To structure this collection, we drew from work on intelligibility types~\cite{lim+dey_assessing_intelligibility2009}, question-driven explanation design~\cite{liao2020}, and the separation of information categories into \textit{data}, \textit{system details}, \textit{usage}, and \textit{context}~\cite{schmude2024information}. We further applied the principles of explanation soundness (fidelity, complexity) and completeness (coverage, density)~\cite{chatti_is_2022, guesmi_interactive_2023, kulesza_principles_2015} by introducing a structure of sub-topics and a hierarchy of explanation levels. This combination of question-driven explanations, levels of detail, and user-controlled selection of information aims to support modularity and interactivity~\cite{cheng_explaining_2019, schmude2023, guesmi_interactive_2023} as well as the adjustment of explanations to users' needs~\cite{conati_toward_2021, shulner-tal_enhancing_2022}. Section~\ref{sec:explanation_design} describes how these principles were realized in the explanation design.

\subsection{Analytical lenses: Understanding and deliberation}

In the following, we introduce understanding and deliberation to serve as the main analytical lenses for this paper. Section~\ref{sec:analysis} then operationalizes them for the evaluation of the explanation design and settings. 

\subsubsection{Individual and collaborative understanding of AI systems}
\label{sec:related_work_understanding}

% Give an idea of what understanding is and how it can be measured, provide reasoning for why we elicit understanding this way, make clear the differences between understanding in groups and individually

Improving understanding of an AI system is the primary purpose of explanations, as understanding is thought to enable assessment (e.g., of a system's fairness)~\cite{langer_what_2021} and action (e.g., contestation)~\cite{henin_beyond_2022} for the system's stakeholders. However, understanding can be defined in numerous ways~\cite{grimm_varieties_2019, baumberger_what_2017, zagzebski_toward_2019, keil2006}. This work draws from research in learning sciences, cognitive sciences, and explainable AI to define understanding as i) connecting and applying information~\cite{grimm_varieties_2019, baumberger_what_2017}, ii) being the attempt to grasp the underlying structure of a phenomenon by way of simplification~\cite{zagzebski_toward_2019}, iii) consisting of several ``facets'' that include both the analytical and the emotional connection to information (explain, interpret, apply, take perspective, empathize, self-reflect)~\cite{wiggins_understanding_2005}, and iv) being a ``working'' mental model that is attained by recognizing and filling gaps until the learner deems it sufficient~\cite{keil2006}. Due to the challenge of defining and measuring understanding~\cite{sato_testing_2019}, recent research has proposed an ``abilities-based'' approach~\cite{speith_conceptualizing_2024}, connecting to comparable operationalizations by the learning sciences~\cite{wiggins_understanding_2005}. We examine understanding by analyzing which facets of understanding participants use to answer the study tasks and make a confident deployment decision (Section~\ref{sec:methods_study_elements}). 

While individual understanding has been the subject of many studies in XAI~\cite{chen2023_machine_explanations, cheng_explaining_2019, wang21, schmude2023}, understanding in group settings has been less explored~\cite{chiang_are_2023}. We thus draw from disciplines that have investigated collaborative understanding: The cognitive sciences have examined distributed cognition (sharing cognitive load) and outsourcing~\cite{keil2006} (delegating understanding) as fruitful mechanisms for collaborative settings, such as the navigation of a ship~\cite{keil_folkscience_2003}. The prerequisite is that groups achieve ``cognitive symbioses with mutually supporting roles''~\cite{keil2006}, i.e., a constructive working dynamic. Similarly, educational psychology has found that peer discussion~\cite{smith_why_2009}, collaborative reasoning~\cite{moshman_collaborative_1998}, and aggregated knowledge~\cite{navajas_aggregated_2018} leads groups to perform better than
individuals on the same tasks. However, whether groups perform well depends on their interactions, which can be described with cognitive and social mechanisms of collaborative success and failure~\cite{nokes-malach_when_2015}. When groups perform worse than individuals, the associated mechanisms include increased memory load and retrieval disruption (losing train of thought). In contrast, members tend to have established common ground and shared task-related information when they perform better. Thus, while it is not clear from the outset if groups are better for learning than one-on-one settings~\cite{bloom_2_1984}, their advantages, such as sharing of cognitive load and exchange of views, likely support finding solutions to complex problems and present a valuable testing ground to deliberate deployment of AI systems.

The field of computer-supported cooperative work (CSCW) has long examined collaborative settings about group composition and interactions~\cite{fiesler_qualitative_2019, convertino_cache_2008, sutcliffe_applying_2005}. However, work in XAI has only begun to consider how explanations for group interactions could be approached, describing that ``many-to-one'' interactions (multiple people interacting with an explanation) will likely differ from ``one-to-one'' interactions due to ``complexities in group dynamics, cognitive bias amplification, trust issues within the group, and group-centric evaluation''~\cite{soton2024}. Lastly, previous work in XAI has examined individual versus group understanding in AI-assisted decision-making but surprisingly found little effect on understanding~\cite{chiang_are_2023}. Following up on these findings, we use a `triangulation of methods'~\cite{carter_use_2014}, as described in Section~\ref{sec:analysis}), to empirically explore and compare the effects of explanations on the understanding processes of AI novices in groups and individual settings. 

\subsubsection{Deliberating on AI systems}
\label{sec:related_work_deliberation}

% Justify why it makes sense to let AI Novices have a say in the deployment of AI, make clear that this format does not exist yet, suggest advantages of using our approach

Deliberation, in the sense of informed reasoning and decision-making, is based on understanding~\cite{definelicht2020} and is key in enabling citizens to debate public sector AI proposals and their potential consequences~\cite{kawakami2024, zuger_ai_2023}. \citet{habermas1991structural} describes deliberation as the exchange of rational-critical arguments on a problem to the end of finding a solution. These rational-critical arguments are grounded in truth or a \textit{shared understanding} of reality, are open for judgment, and can be defended. This connection between shared understanding and deliberation is central to our examination of explanations' effects. Deliberation takes place in many areas that shape politics and life in society~\cite{lupia_by_2023}. Examples include public referendums that let inhabitants vote on jurisdictional changes (such as Swiss federal and state laws~\cite{switzerland_referendums}), citizen forums addressing matters of public importance (such as water supply in California~\cite{innes_collaborative_2003}), and community-based grassroots formats where citizens support each other (such as the right to repair movement~\cite{collins_right_2024}). These settings have in common that they involve ``social entities made up of people who are in one way or another engaged with their environment'' and who use deliberation and productive conflict to negotiate and change policy issues~\cite{hajer_deliberative_2003}. While these participation formats are not perfect and potentially incur cognitive biases such as \textit{groupthink}\footnote{Prioritizing group harmony over real argumentation and discussion.}~\cite{soton2024, janis1971groupthink, baron_so_2005}, they create spaces where the general public can gather, discuss, form opinions, and decide on public interests. We argue that AI systems in public institutions constitute such public interests, encapsulated in the term \textit{public AI}.~\citet{zuger_ai_2023} employ the term to make explicit that AI systems in public institutions must fulfill obligations to prove their benefit. These obligations include being justifiable, equal, open to validation, technically secure, and the result of a \textit{deliberation or co-design process}. Identifying formats supporting this deliberation on public AI systems is an open research challenge. Prior work has investigated how 'mini-publics'~\cite{fung2003} can be used to support the co-design of algorithmic policy~\cite{lee_webuildai_2019} and procedural justice in algorithmic resource allocation~\cite{lee2019procedural}. HCI research has further shown that participatory formats can connect communities and institutions in public service transformation~\cite{crivellaro_2019}. And in XAI, studies showed that group discussions can facilitate the critical analysis of an AI system's recommendations and that supplying information on both pros and cons of an AI's recommendation lead to more frequent and more productive group deliberation~\cite{chiang_are_2023}. 

However, settings that allow participants to deliberate in person on the deployment of high-stakes public AI are underexplored. We aim to address this gap by implementing mini-publics as focus groups with three different compositions (domain experts, decision subjects, and members of the general public\footnote{Participants who were neither directly affected as job-seekers nor were potential users of the system.}), thus including stakeholders of different backgrounds and degrees of involvement. We further compare group deliberation processes with those in single interviews, which can be described as ``internal deliberation''~\cite{mercier_reasoning_2012}. On this basis, we aim to provide insight into suitable explanation designs and social formats to support deliberation on public AI systems. Section~\ref{sec:analysis} describes the concrete analysis approach to this end.

\section{Methods}
\label{sec:method}

In this section, we describe our methods and study procedure. We conducted a task-based semi-structured interview study with 43 participants (Section~\ref{sec:participants}), structured into 8 focus groups with 3--5 participants each and 12 single interviews (Section~\ref{sec:study_procedure}). % to compare whether group or single settings impacted how people interacted with the explanations and whether they affected self-reported understanding and decision confidence. 
Participants were presented with the study's employment prediction use case (Section~\ref{sec:use_case}) and a collection of explanations about this system (Section~\ref{sec:explanation_design}) before solving four tasks and deciding about the system's deployment (Section~\ref{sec:methods_study_elements}). The study closed with an interview, lasting 90--120 minutes for focus groups and 60 minutes for single interviews. We analyzed individual and collective interactions with the explanations, self-reports, and deliberation processes (Section~\ref{sec:analysis}). The university’s research ethics committee approved this study. 

\subsection{Use case: The \textit{AMS} employment prediction algorithm} 
\label{sec:use_case}

\subsubsection{Description} The \amsalgorithm\footnote{AMS stands for the Public Employment Agency (Arbeitsmarktservice).} is a system developed to calculate the employability of job-seekers in Austria. It was created by a private company for the Austrian Public Employment Agency between 2015 and 2021 but was never used as a live system and put on hold in 2021 due to privacy objections~\cite{allhutter_bericht_ams-algorithmus_2020}. The system uses a logistic regression model trained on historical data to predict job-seekers' employment chances based on demographic features (such as age, education, nationality, etc.) and work history. The outputs are a short-term and long-term employment score for each job-seeker~\cite{gamper_assistenzsystem_2020}. These scores would serve as recommendations for the job-seekers' counselors at the employment agency to assist in deciding about suitable support measures. Counselors could overwrite the system's predicted scores of job-seekers but would need to give a reason for doing so~\cite{allhutter_bericht_ams-algorithmus_2020, Holl2018}. More information is provided in the appendix.

\subsubsection{Choice of use case} Algorithmic tools that assist in assessing job-seekers and resource allocation have been deployed in various countries, including Germany~\cite{agentur_für_arbeit_2021}, Austria~\cite{allhutter_bericht_ams-algorithmus_2020}, Poland~\cite{niklas2015}, and the Netherlands~\cite{DESIERE_STRUYVEN_2021}. However, the introductions of these applications also repeatedly led to sociotechnical conflicts~\cite{scott_algorithmic_2022}. The deployment of the \amsalgorithm was motivated by three overarching goals: a) increasing consultation efficiency, b) increasing support measure effectiveness, and c) reducing arbitrariness~\cite{gamper_assistenzsystem_2020}. Detailed reports warned that counselors might over-rely on the algorithm or hesitate to overrule its suggestions~\cite{allhutter_bericht_ams-algorithmus_2020}. Further, the algorithm's model and underlying data structure were predicted to discriminate against marginalized groups, who would lack the option to contest the system itself~\cite{lopez_reinforcing_2019}. Transparency and ongoing scrutiny of the algorithm were listed as necessary measures to prevent these risks~\cite{allhutter_bericht_ams-algorithmus_2020}. As the \amsalgorithm represents a larger class of algorithmic decision-making systems that spark public debate around their deployment in public institutions~\cite{raji_fallacy_2022}, it exemplifies how AI systems become matters of public interest and presents a suitable use case for our study.

\subsection{Explanation design}
\label{sec:explanation_design}

\subsubsection{Description} The explanation design comprised 36 question-answer pairs about the \amsalgorithm. Each question belonged to one of four categories, \textit{data} (format, content, limitations), \textit{system details} (features, model, examples), \textit{usage} (operation by and interaction with users), and \textit{context} (intention of deployment, target group, responsible actors). Each category was further divided into topics with three levels of increasing detail (base level, level 2, level 3). Every explanation was printed on an A5 paper sheet and contained a question (e.g., "Who operates the system?") answered with a brief text or image (cf. \autoref{fig:explanation_examples}). Participants first received an explanation overview (\autoref{fig:explanation_overview}) and the four base explanations and could request levels 2 and 3 at any time during the explanation phase (as depicted in~\autoref{fig:study_procedure}). The explanations were presented in an analog paper format that allowed participants to interact with them physically and that facilitated social interactions, such as sorting, exchanging, pointing, and reading to each other. The full collection of explanations is depicted in the appendix.

\subsubsection{Design foundations} The explanation design was intended to allow the users to explore information in a flexible and self-directed manner. To this end, the design used a \textit{modular structure}, meaning that the explanations were divided into four information categories, which were again subdivided into topics and three levels of detail (base level, level 1, level 2). The four information categories, \textit{data}, \textit{system details}, \textit{usage}, and \textit{context} were based on research on AI novices' information needs and covered both technical and sociotechnical system aspects~\cite{schmude2024information}. The subdivision of explanation categories into topics and levels of detail organized this broad supply of information while accommodating different needs of information completeness and soundness~\cite{kulesza_too_2013} and introducing a degree of personalization~\cite{chatti_is_2022}. The goal of the modular design was thus to create explanations that offered information on every aspect of the AI system, from which participants could select the most relevant according to their information needs and preferences. It further aimed to avoid limitations of ``groupware'' systems~\cite{mandviwalla_1994} by supporting both individual and collaborative interaction, providing multiple user perspectives (e.g., user, decision subject), and synchronizing interaction with the material. 

\begin{figure}[H]
    \centering
    \includegraphics[width=\textwidth, bb=0 0 1200 442]{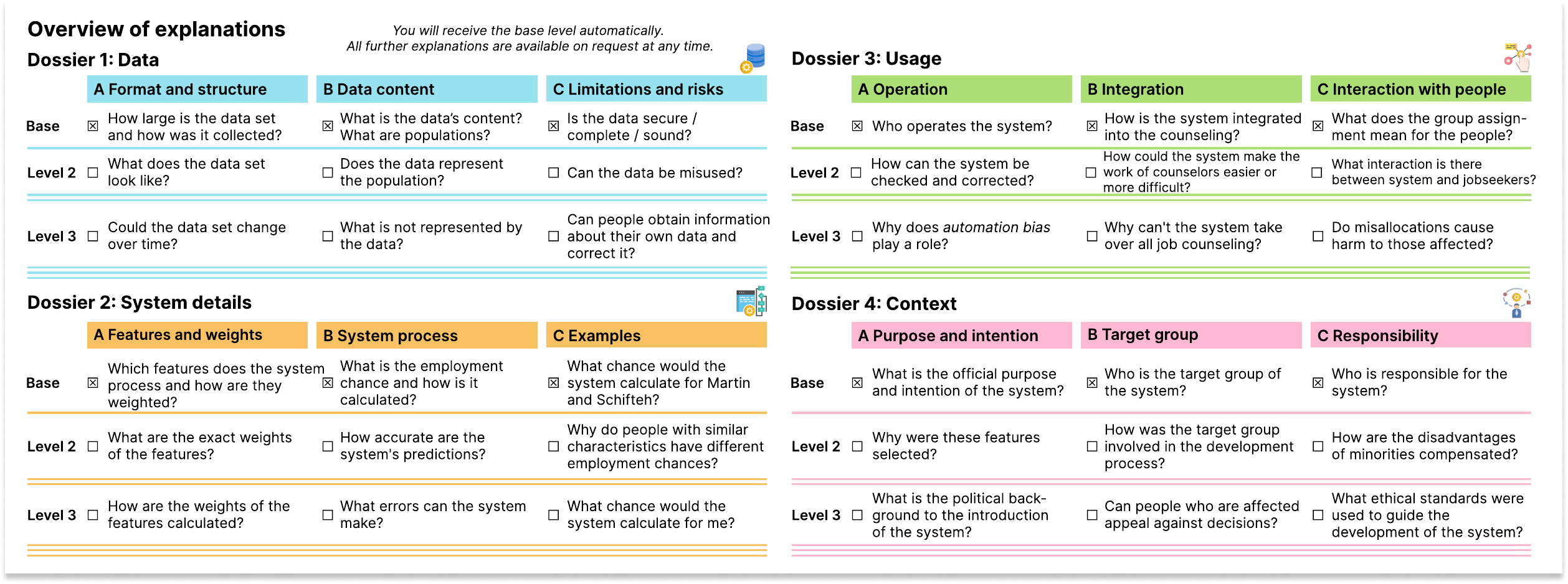}
    \caption[Explanation overview]{\textbf{Overview of explanations.} Explanations were designed as a collection of 36 question-answer pairs. The questions were assigned to 4 categories, \textit{data}, \textit{system details}, \textit{usage}, and \textit{context}, each containing 9 questions. Participants received the base explanations at the beginning of the explanation phase, as indicated by the ticked boxes, and could request all other explanations at any time during the explanation phase using this overview.}
    \label{fig:explanation_overview}
\end{figure}

\begin{figure}[H]
    \centering
    \includegraphics[width=380pt, bb=0 0 1226 865]{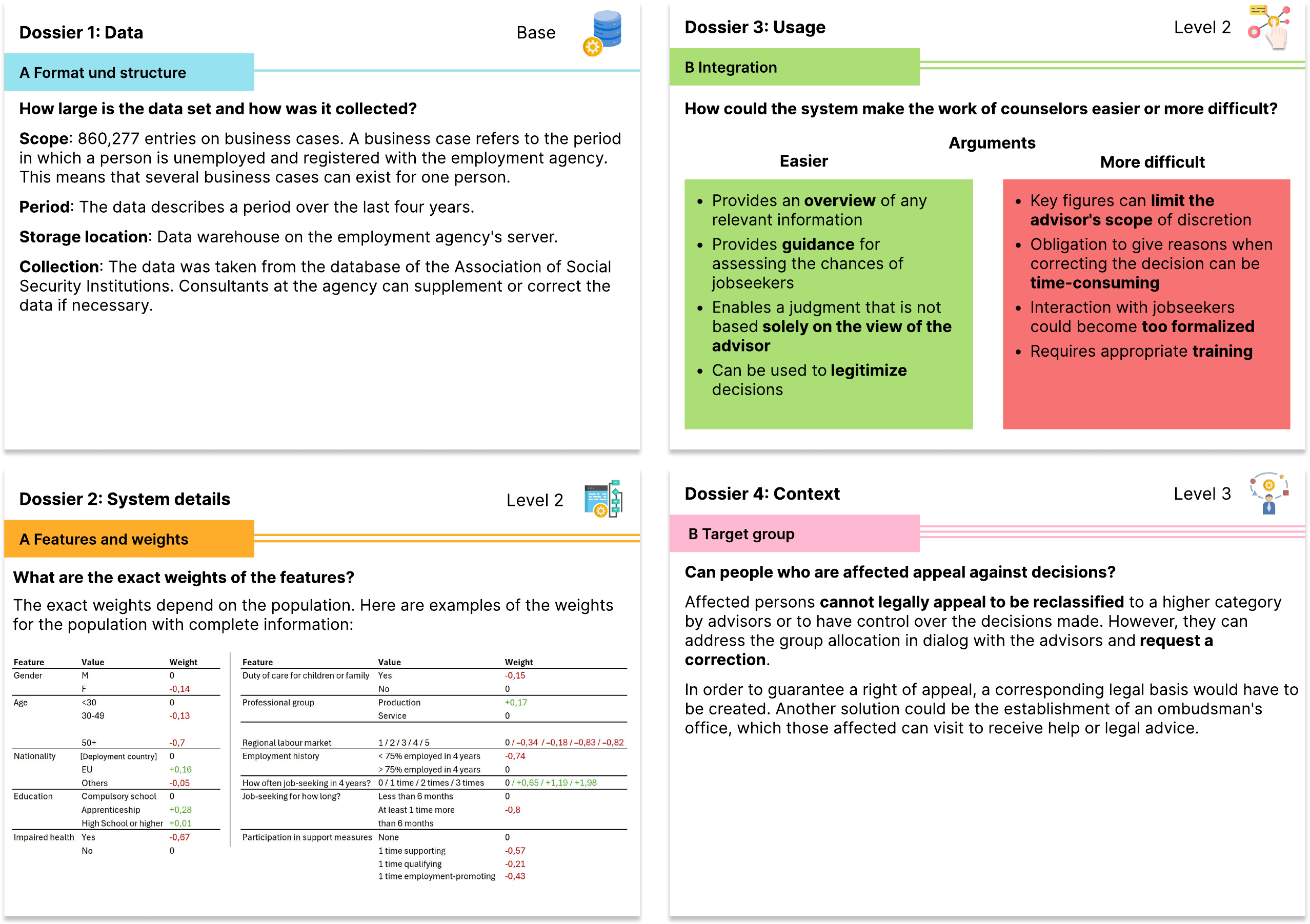}
    \caption[Explanation examples]{\textbf{Four explanation examples.} Examples for explanations in the categories \textit{data}, \textit{system details}, \textit{usage}, and \textit{context}. Each question was printed on a sheet of A5 paper with a short answer to the question. Answers could be fully textual or complemented with visual elements like charts or colored shapes. Each category was given a different color and icon to facilitate navigation.}
    \label{fig:explanation_examples}
\end{figure}

The question-answer style was motivated by explanation design research~\cite{liao2020, liao2021questiondriven} and was intended to improve user engagement and understanding by matching their thought processes. For example, when users learn that the system uses features to calculate job-seekers' employment scores, they might ask what the exact weights of these features are and how they are calculated. This corresponds to the three levels of detail in topic A of \textit{system details} (\autoref{fig:explanation_overview}). The explanations further used different explanation methods~\cite{speith_review_2022} such as feature importance, local and global explanations, examples, counterfactuals, and argumentative approaches (\autoref{fig:explanation_examples}). Information was presented in different formats but mostly relied on textual information and used highlighting, colors, and illustrations to emphasize key points. 

\subsection{Study procedure}
\label{sec:study_procedure}

\begin{figure*}[h]
    \centering
    \includegraphics[width=\textwidth, bb=0 0 1799 546]{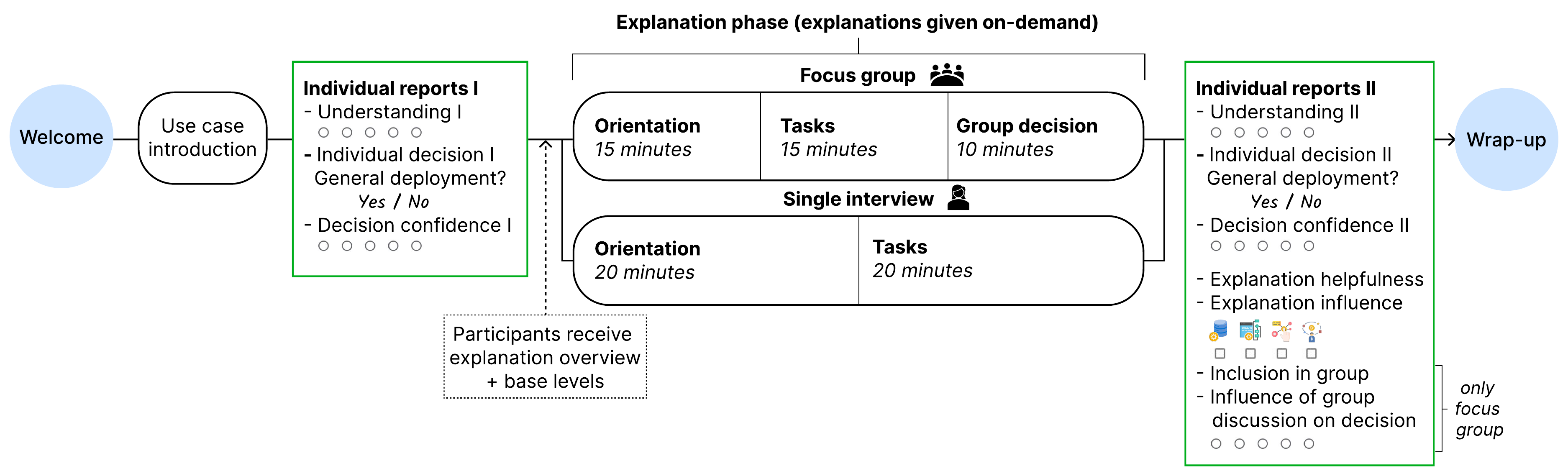}
    \caption[Study procedure]{\textbf{Overview of the study procedure.} Focus groups and single interviews differed only in the explanation phase and the questions for the second individual reports.}
    \label{fig:study_procedure}
\end{figure*}

We describe the procedure in the focus group and single interview setting (depicted in~\autoref{fig:study_procedure}). Like previous work in XAI~\cite{chiang_are_2023}, we conducted individual and group settings to compare how the social setting would affect participants' understanding and deliberation processes. 

\subsubsection{Focus group procedure.} Throughout the study, participants sat together with the investigator and could freely interact with each other. They first completed consent forms and questionnaires about demographics and knowledge about employment (domain knowledge) and AI systems (technical literacy). A round of introductions followed, where each group member stated their name and last interaction with AI to break the ice. The investigator then explained the study procedure and distributed a mock newspaper article introducing the AI use case (included in the appendix). Participants indicated their understanding, deployment decision, and decision confidence for the first time (\ref{sec:methods_study_elements}). The study's explanation phase followed (including orientation, task, and decision phase), throughout which participants received and kept access to all explanations. In the beginning, the group received an overview of the explanations and all base level explanations, all other explanations could be requested at any time. After 15 minutes, participants received task sheets and had another 15 minutes to complete them, deciding independently whether they wanted to work together or individually. Finally, the group had 10 minutes to make a joint deployment decision (yes/no, with conditions allowed). A second round of individual reports followed, described in the next section, and then the investigator concluded the study. Focus group studies took around 90 minutes.

\subsubsection{Single interview procedure.} The majority of the study procedure remained the same in single interviews. However, the explanation phase did not include a group decision phase; instead, the orientation and task phases were prolonged to 20 minutes each to provide individuals with the same total time as focus groups to interact with the explanations and individual reports about the group setting were omitted. The individual study setting was included to analyze how understanding processes changed depending on whether participants worked in a group or alone and whether the explanation design would support both settings. Single interviews took around 60 minutes. 

\subsection{Study elements}
\label{sec:methods_study_elements}

This section provides descriptions and motivation for the study elements: the introduction of the use case, the explanation phase comprising orientation, tasks, and the group decision, and participants' individual reports.

\subsubsection{Use case introduction.} Participants received initial information about the \amsalgorithm in the form of a mock newspaper article inspired by an Austrian newspaper publication from 2019~\cite{standard_ams_2018}. The article provided key information about the system's basic workings, goals, and deployment context and featured the opinions of employers and employee associations about its merits and risks. The presentation format was chosen to provide an introductory summary of the AI system using a familiar layout and non-technical language while highlighting both the pros and cons of the system's deployment. Thus, The article served as a basic introduction to the use case, which aimed to approximate the amount of information participants might receive through the media. This way, participants received a baseline of information with which to assess their initial understanding, deployment decision, and decision confidence. Further, this introduction served to outline relevant aspects that could be explored using the explanations.   

\subsubsection{Explanation phase (orientation, tasks, and group decision).} For orientation, participants received the explanation overview (\autoref{fig:explanation_overview}) and base-level explanations (the first row of explanations in each category) and had 15 minutes (focus groups) or 20 minutes (single interviews) to become familiar with the structure and explore topics of interest. Explanations were provided as single A5 sheets to promote the physical sharing and exchanging of explanations. Participants could freely decide which explanations to request and read, and whether to share and discuss information with others. 

For the tasks, participants received the case of Mr. Harald G.\footnote{The case example was inspired by~\citet{allhutter_bericht_ams-algorithmus_2020} and adapted to this study, as depicted in the appendix.}, a fictional job-seeker with a brief backstory and a list of features. Participants had 15 minutes (focus groups) or 20 minutes (single interviews) to solve four tasks pertaining to this case. All questions could be answered with information from explanations in different categories and levels of detail. Whereas tasks 1, 3, and 4 required locating information, task 2 could be solved in two ways (aside from guessing): by either giving an estimate based on the rough weightings in the \textit{system details} base explanations or calculating the precise employment score. Participants could access and request all explanations and discuss possible solutions. These were the four tasks (correct answers underlined):

\begin{quote}
    \textbf{Task 1}: Can Harald change the data stored about him (e.g., to correct it)? (\underline{yes} / no) \\[1mm] 
    \textbf{Task 2}: In which group of employment chance does the system categorize Harald? \\ (high (>66\%) / medium (<66\% \& >25\%) / \underline{low (<25\%)}) \\[1mm]
    \textbf{Task 3}: Which support measures will Harald receive? \\ (qualifying measures, such as courses and training / \underline{stabilization and increased supervision} / none) \\[1mm]
    \textbf{Task 4}: Can Harald appeal against this decision? (yes / \underline{no})
\end{quote}

\noindent

For the group decision, focus group participants had 10 minutes to discuss the system's deployment and were asked to collectively decide whether to accept or reject it. This was meant to simulate a small referendum in which each participant's voice counted for the final outcome. If no consensual decision was reached in time, participants were asked how the situation should be resolved (e.g., majority vote). They were further made aware of the option to state conditions for the system's deployment. 

\subsubsection{Individual reports I and II} Participants were asked for individual reports before and after the explanation phase. At both points, participants reported understanding (5-point scale), deployment decision (yes / no), and decision confidence (5-point scale) to examine the effect of the explanation phase. In report II, participants also reported the explanation categories that were most helpful for their understanding (multiple choice) and most influential to their decisions (multiple choice); focus group participants additionally reported perceived inclusion in the group and the discussions' influence on their decision (5-point scales). During report II, the investigator asked participants interview questions about their interaction with the explanations, understanding processes, prioritized information, and situational aspects. The list of interview questions is included in the appendix. 

\begin{figure*}[h]
    \centering
    \includegraphics[width=350pt]{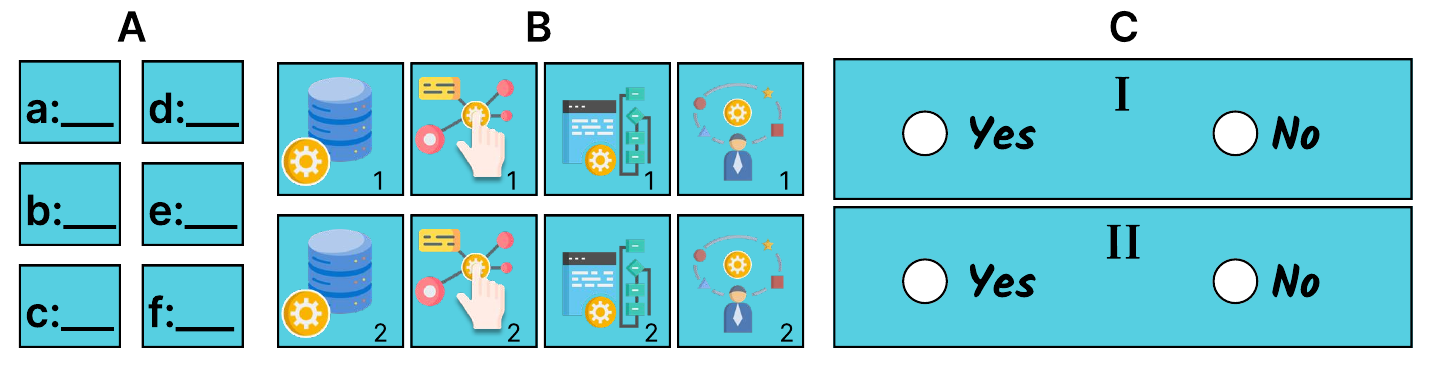}
    \caption[Self-report material]{\textbf{Material for individual reports of participants.} Participants received the materials for individual reports on laminated paper slips in different colors (blue, yellow, red, green, and grey) and used them to answer questions individually. Slips that were numbered with letters a to f (A) served as 5-point scales for understanding, confidence, inclusion in group, and influence of group discussion and were answered by writing a number (1--5); slips with icons and a corresponding textual description (B) served as selection of the most helpful and influential explanation categories and were answered by selecting any number of icons; slips with decisions (C) served as voting ballots for deployment decisions and were answered by ticking yes or no.}
    \label{fig:voting_material}
\end{figure*}

To prevent influence between participants' reports, individual reports in focus groups were conducted anonymously and re-assigned by the study examiner using a color-coded reporting system (\autoref{fig:voting_material}). Each participant was assigned a color and received the material for all ten individual reports. Participants took the corresponding paper slip for each report, wrote their answer, and threw it in a gathering container that hid it from view. The gathered reports were then collected and recorded by the investigator at the end of the study. 

\subsection{Analysis}
\label{sec:analysis}

All focus groups and interviews were audio-recorded and transcribed. These transcripts provided the data basis for the thematic analysis. Participants' individual reports, task solutions, decisions, and the investigators' field notes provided further data for within- and between-subject comparisons of understanding and decision-making. 

\subsubsection{Thematic analysis.} For both research questions, we conducted thematic analysis~\cite{braun_using_2006} of participants' articulations to develop a qualitative account of their understanding and decision-making processes. To this end, we developed inductive codes in the first pass and refined them in the second pass on the transcriptions. The resulting inductive code base was structured along the overarching categories of understanding, deliberation (decision-making processes, arguments), opinions (e.g., about AI and policy choices), and experiences (e.g., anecdotes and lived situations). The full code-book is provided in the appendix. We further highlight that while the quantitative items in participants' self-reports serve to characterize the diversity of participants' perceptions and facilitate qualitative exploration~\cite{Weiss1995}, they are not intended to invoke ``inference [...] of greater generality''~\cite{maxwell2010} nor impose a mental model based on variance theory~\cite{patton1990}.

\subsubsection{RQ1-Explanations.} RQ1 focuses on the explanations' impact on participants' understanding of the study's use case and the differences between individual and group settings. We employ a triangulation approach~\cite{carter_use_2014} by using three ways of analyzing the explanations' effect on participant understanding. Firstly, comparing participants' individual reports before and after the explanation phase was a subjective indicator of changes in their understanding and decision confidence. Secondly, participants' answers to the four study tasks indicated their factual or testable understanding. Lastly, participants' verbal reports during and after the explanation phase were used to analyze their understanding processes and barriers thematically. In a deductive analysis, we further compared their interactions to mechanisms of ``collaborative success and failure''~\cite{nokes-malach_when_2015} and the ``six facets of understanding''~\cite{wiggins_understanding_2005}. With this three-part combination, we examined participants' subjective understanding, their information gain, and the cognitive processes of their understanding. This choice was motivated by educational psychology research outlining that understanding cannot only be elicited through test questionnaires~\cite{sato_testing_2019} but involves emotional~\cite{wiggins_understanding_2005} and meta-cognitive processes~\cite{veenman_metacognition_2006} that are equally important. Our focus on understanding is motivated by previous XAI research, which highlights the importance of understanding in decision-making processes~\cite{langer_what_2021, shin_algorithms_2023, hoffman_explainable_2023}.

\subsubsection{RQ2-Deliberation.} RQ2 focuses on how participants formed opinions about the \amsalgorithm, weighed the pros and cons of its deployment, and settled on a deployment decision. To this end, we compare participants' decision confidence before and after the explanation phase. We further conduct an inductive and deductive thematic analysis of participants' interactions in both settings to connect them to the ``elements of deliberation''~\cite{stromer-galley_measuring_2007} -- a set of characteristics that outline deliberation processes. Based on this, we analyze when participants used arguments (grounded, defensible positions), opinions (personal judgments on things, values, states), and personal experiences~\cite{stromer-galley_measuring_2007, mercier_reasoning_2012} to consider the system's deployment. For single interviews, we examine participants' responses to interview questions during the study to examine their reasoning process and ``internal deliberation''~\cite{mercier_reasoning_2012}. Lastly, to account for one of the most prevalent cognitive biases in group settings, we examine focus groups for occurrences of groupthink~\cite{janis1971groupthink, baron_so_2005} -- an effect that sets in when concurrence-seeking in groups overrides realistic argumentation and discussion.

\subsection{Participants}
\label{sec:participants}

\subsubsection{Recruitment.} \autoref{fig:group_articipant_table} and \ref{fig:interview_participant_table} provide an overview of the study participants. Participants were recruited through cooperation with civil society organizations, an employment agency, public calls for participation, and the authors' extended network. For focus groups, the authors contacted staff from these organizations known from previous studies or used channels of general inquiry to describe the study and invite participation. Interested organizations all offered to support the recruitment process by coordinating with the authors on selecting and inviting potential participants and finding a place and time to conduct the studies. Groups A, B, C, E, F, G, and H were organized this way. Group D was recruited through the authors' network and was equally composed of people who had previously been job-seeking. For individual studies, participants were recruited using the same channels, and calls for participation were further advertised on screens and information boards throughout different city districts. All studies were conducted in person in office or public spaces. Participants were compensated with 30€ for participation in focus groups (90--120 minutes) and 20€ for single interviews (60 minutes). Our approach for organizing, composing, and moderating focus groups was informed by~\citet{KruegerRichardA2004Fg:a}. Concerning the participant sample size, we are guided by research on qualitative methods, which suggests that the number of participants should be determined by code and meaning saturation~\cite{hennink_code_2017}.

\subsubsection{Recruitment criteria.} Participants were required to be of full legal age and AI novices, i.e., to have no technical knowledge or expertise about machine learning systems as described in Section~\ref{sec:related_work_ai_novices}. These criteria were screened in a pre-questionnaire before invitation to the study using two questions: ``How would you rate your knowledge of algorithms?'' and ``How would you rate your knowledge of artificial intelligence (AI)?''. Each question could be answered on a scale corresponding to ``no knowledge at all'' to ``professional and detailed knowledge''. Here, the first question elicited technical knowledge, as participants familiar with AI tools might have rated their AI expertise as high but were unlikely to know about algorithms without a strong technical interest or background.

\subsubsection{Group composition.} Participants were further selected to be representatives of one of three roles: domain experts, decision subjects, or members of the general public. We define domain experts as people who are competent in the field that the AI system is used in, such as job counselors or advisors (groups B, C, G, and H). We define decision subjects as people who would potentially be impacted by the system's decision, such as job-seekers and people who had previously been job-seeking (groups E and F). All remaining participants are considered members of the general public and were included to test changes in explanation effects and participants' perceptions (groups A and D). The study was conducted with separate participants in three pilot groups to test and refine the explanation design and study procedure.    

\begin{table}[H]
\centering
\caption{\textbf{Details on the study participants in the focus groups.}}
\resizebox{\textwidth}{!}{%
\begin{tabular}{lllll|lllll}
\hline
Group   & ID & Age & Education         & Occupation          & Group   & ID & Age & Education         & Occupation           \\ \hline
Group A & A1 & 63  & University        & Retired             & Group F & F1 & 48  & A-levels          & Job-seeking          \\
       & A2 & 69  & Secondary school  & Retired             &        & F2 & 35  & n/a               & Job-seeking          \\
        & A3 & 63  & Vocational school & Retired             &         & F3 & 49  & A-levels          & Job-seeking          \\
        & A4 & 70  & Vocational school & Retired             &         & F4 & 50  & Vocational school & Job-seeking          \\ \cline{1-5}
Group B & B1 & 46  & University        & Social counselor    &         & F5 & 48  & A-levels          & Job-seeking          \\ \cline{6-10} 
       & B2 & 76  & A-levels          & Retired             & Group G & G1 & 37  & University        & Executive staff      \\
        & B3 & 46  & University        & Social counselor    &        & G2 & 49  & University        & GDPR officer        \\
        & B4 & 70  & A-levels          & Retired             &         & G3 & 44  & Secondary school  & Training counselor   \\ \cline{1-5}
Group C & C1 & 60  & Apprenticeship    & Personnel counselor &         & G4 & 58  & University        & Executive staff      \\ \cline{6-10} 
       & C2 & 60  & University        & Personnel counselor & Group H & H1 & 37  & University        & Team lead            \\
        & C3 & 51  & Apprenticeship    & Job trainer             &        & H2 & 56  & Apprenticeship    & Job trainer              \\ \cline{1-5}
Group D & D1 & 65  & University        & Business consultant &         & H3 & 45  & University        & Job trainer              \\
       & D2 & 53  & University        & Retired             &         & H4 & 43  & University        & Job trainer              \\
        & D3 & 52  & University        & Business consultant &         & H5 & 60  & University        & Administrative staff \\ \hline
Group E & E1 & 36  & University        & Graphic designer    &         &    &     &                   &                      \\
       & E2 & 32  & Apprenticeship    & Job-seeking         &         &    &     &                   &                      \\
        & E3 & 40  & Apprenticeship    & Job-seeking         &         &    &     &                   &                      \\ \cline{1-5}     
\end{tabular}}
\label{fig:group_articipant_table}
\end{table}

\begin{table}[H]
\centering
\caption{\textbf{Details on the study participants in the single interviews.}}
\resizebox{350pt}{!}{%
\begin{tabular}{llll|llll}
\hline
ID & Age & Education  & Occupation           & ID  & Age & Education  & Occupation               \\ \hline
S1 & 74  & University & Retired              & S7  & 40  & University & Job trainer                  \\
S2 & 29  & A-levels   & Nurse                & S8  & 43  & University & Rehabilitation counselor \\
S3 & 28  & University & Social counselor     & S9  & 44  & University & Social center manager    \\
S4 & 29  & University & Doctoral student     & S10 &  52 & University & Rehabilitation counselor    \\
S5 & 37  & University & Administrative staff & S11 &  59   & University           &  Social center manager                        \\
S6 & 28  & University & Job-seeking          & S12 &  39   & University           &  Education program manager                         \\ \hline
\end{tabular}}
\label{fig:interview_participant_table}
\end{table}

\newpage

\section{Results}
\label{sec:results}
% Why are our explanations good to understand AI?
% What has been understood that has not been before?

In this section, we present our results as answers to our research questions: How question-driven, modular explanations\footnote{We note again that with ``explanation'' we mean a question and answer pair and with ``explanations'' we mean the collection of all 36 explanations (Section~\ref{sec:method}).} support understanding in individual and group settings (RQ1, Section~\ref{sec:rq1}) and how AI novices used explanations to form opinions and decide about the system's deployment (RQ2, Section~\ref{sec:rq2}). Participant labels denote the study setting (focus group: A--H / single interviews: S) and the participant ID, as listed in Tables~\ref{fig:group_articipant_table}+\ref{fig:interview_participant_table}. To distinguish themes in the analysis, \textit{inductive themes} are italicized, while `deductive themes' are put in quotes.

\subsection{RQ1-Explanations: How does a question-driven, modular explanation design support AI novices’ understanding in groups and individual settings?}
\label{sec:rq1}

To examine how AI novices used the explanations to understand the study's use case, we analyzed their self-reports, articulations, and interactions in both settings. We found that each setting supported different aspects of understanding, suggesting a trade-off. We first describe how the explanations contributed to \textit{shared understanding} and `collaborative success' in groups (\ref{sec:results_groups_benefits}) and continue with the explanations' role in instances of `collaborative failure'~\cite{nokes-malach_when_2015} (\ref{sec:results_groups_drawbacks}), summarized in~\autoref{fig:findings_group_understanding}. We then describe individuals' interactions with the explanations (\ref{sec:results_individuals}), participants' feedback on the explanation design (\ref{sec:results_explanation_design}), and summarize the benefits and drawbacks of both settings for XAI (\ref{sec:rq1_summary}).
% In the study, participants received a collection of 36 explanations to solve four study tasks and decide about the deployment of a public AI system (cf.\ Section~\ref{sec:method}). 

\subsubsection{Groups' benefits: Shared understanding and increased engagement.}
\label{sec:results_groups_benefits}

In the best cases, groups leveraged the modular explanation structure to use distributed cognition~\cite{keil_folkscience_2003}, meaning that participants processed information in parallel and then combined it. We use the term \textit{shared understanding} to capture interactions that realized distributed cognition. Examples of such interactions included \textit{locating information together}, \textit{sharing information with others}, \textit{discussing interpretations}, \textit{debating task solutions}, and \textit{querying and explaining} (a question by a group participant invites other participants to contribute). The explanations only afforded this set of interactions to groups, as they required social interaction with other participants. For example, in Group C, participant C1 read the first study task aloud and asked for input (\textit{querying}), after which the group discussed solutions (\textit{explaining}):

\begin{quote}
    \colorbox{GroupGreen}{\textbf{C1}} \textit{Can Harald change the data stored about him? Yes, he can certainly change it, can't he?} [...] \\
    \colorbox{GroupRed}{\textbf{C3}} \textit{Which stored data, the one down there?} [points at Harald's demographic features] \\
    \colorbox{GroupGreen}{\textbf{C1}} \textit{Yes, just that.}\\
    \colorbox{GroupRed}{\textbf{C3}} \textit{49 -- no, male -- no. The apprenticeship -- no, Austria -- he can still change that. Duty of care -- he could get married or have children. He could change his service sector. He could change his career. Impairment...} \\
    \colorbox{GroupGreen}{\textbf{C1}} \textit{Well, what is meant by `change'? When he enters the data, he can change the data. He doesn't have to specify the knee problem.} [...] \\
    % \textcolor{GroupBlue}{\textbf{C2}} \textit{He can do that too.} [...] \\
    \colorbox{GroupRed}{\textbf{C3}} \textit{So he can change it.}\\
    \colorbox{GroupBlue}{\textbf{C2}} \textit{Yes.}
\end{quote}

Interactions such as \textit{querying and explaining} and \textit{discussing interpretations} rely on collaboration between participants to `share working memory resources', `complement others' knowledge', `re-expose information', and `correct errors'. These aspects of collaboration are described as cognitive mechanisms of `collaborative success'~\cite{nokes-malach_when_2015} and provide groups with multiple ways to tackle explanations. For example, participants tended to work through information about \textit{usage} and \textit{context} alone or in pairs but raised explanations with the group when they were difficult or piqued their interest. This was often the case with explanations about \textit{system details}, which included the most numerical information but also were an important key to understanding the system's calculations. We describe the process of using other's understanding to close gaps in one's own as \textit{outsourcing}~\cite{keil2006}. As understanding AI systems involves interacting with a variety of different information categories (e.g., technical, political, social), outsourcing provides a way to hand information to the team member most competent in this category. For example, in Group B, B2 expressed their appreciation for B3's help in solving study task 2: ``\textit{It was a math problem. You [B3] filtered it out well. It was very analytical. With your help, we were able to recognize these weak points.}'' In contrast to \textit{querying and explaining}, which participants used to invite input or spark conversation, \textit{outsourcing} was thus used for the active delegation of an impeded understanding process. 

Whether groups used this collaboration depended on participants' relationships and the group's social dynamics. The explanations also served to support these `social mechanisms' of collaboration, by drawing the group's attention to certain aspects of the AI system and encouraging them to share their experiences and opinions. We present an excerpt from Group G as an illustration. Here, participant G4 \textit{shared an explanation} that documented the algorithm’s impact on two job-seekers (`joint management of attention'), which prompted G2 and G3 to \textit{discuss interpretations} (`increased engagement'). This interaction established `common ground' that the group later used for deliberation: 

\begin{quote}
    \colorbox{GroupYellow}{\textbf{G4}} \textit{That's bad, the two of them. Look, ``What chances would the system calculate for Martin and Schifteh?''} \\
    \colorbox{GroupBlue}{\textbf{G2}} \textit{Schifteh is probably worse off, isn't she?} \\
    \colorbox{GroupYellow}{\textbf{G4}} \textit{Schifteh has a 30\% chance of employment and Martin 52\%, even though Schifteh has a degree and would be working in the IT sector. And Martin has compulsory schooling and works in the cleaning sector. Martin's chances of employment are almost twice as high as Schifteh's.}  [...] \\
    \colorbox{GroupRed}{\textbf{G3}} \textit{I think that's a bit weird.} [...] \textit{Because if she can speak English very well and has the specialist knowledge that our IT sector needs...} \\
    \colorbox{GroupYellow}{\textbf{G4}} \textit{She even gets two minuses for living in Favoriten [a city district].} [...] \\
    \colorbox{GroupGreen}{\textbf{G1}} \textit{Yes, and here you have it in writing, I'll have to look at that too.} 
\end{quote}

A later excerpt in Section~\ref{sec:results_delib_disagreement} further shows that explanations also led participants to `negotiate multiple perspectives'. Addressing both sides of collaboration, cognitive and social, here is an important goal when supporting group understanding. It can be assumed that when explanations succeed in doing both, they can provide participants with more comprehensive and more complex information than individual settings. In line with this, previous work in HCI has found that group interaction boosts task performance compared to individual settings~\cite{karadzhov_delidata_2023}. However, in our study, individuals surprisingly performed better in the study tasks than groups, as discussed in Section~\ref{sec:results_individuals}. Even so, we argue that the set of interactions (\textit{shared understanding}) enabled through the combination of our explanation design and the group setting presents important pathways to help AI novices understand algorithmic systems. These interactions can be especially useful when group members have different domain expertise and information needs, as they can use complementary knowledge, memory, and perspectives to make sense of information. At the same time, as noted earlier, these interactions are partly dependent on the group dynamic. Positive interactions like those described were especially frequent in Groups G and H (job counselors and trainers), where participants knew and trusted each other. In contrast, the next subsection describes instances where groups encountered challenges in understanding, illustrating the importance of social mechanisms.  

\subsubsection{Groups' drawbacks: Process loss and susceptibility to social dynamics}
\label{sec:results_groups_drawbacks}

In some of the focus groups, participants lost track of information, forgot their train of thought, or abandoned understanding altogether. We summarize these effects under the term \textit{impeded understanding} and its final result as \textit{abandoned understanding}. We found that \textit{impeded understanding} occurred due to \textit{explanation design flaws} and co-occurred with adverse social dynamics, resulting in `process loss' (groups falling short of their potential performance~\cite{kerr_group_2004}). For some participants, the benefits of the explanations' modular structure turned into disadvantages when it hampered them in navigating and retrieving information. Such impeded interactions included \textit{cumbersome information uptake}, being \textit{overwhelmed by information}, and \textit{relying on intuition over information}. Further, for some participants, the group setting contributed to these impediments. For example, participant B3 stated that ``\textit{For me, it doesn't make sense to [...] split up [the explanations], and everyone reads a part, that's actually not enough.}'' As before, these impediments can be connected to cognitive mechanisms of `collaborative failure'~\cite{nokes-malach_when_2015}. When groups had difficulties in interacting with the explanations, they also incurred `memory coordination cost' (increased cognitive load) and `retrieval strategy disruption' (losing train of thought). We illustrate these mechanisms with an exchange from Group G. Although this group was composed of participants with university education, it did not succeed in calculating the employment chance for study task 2, in contrast to single interview participants with the same education.

\begin{quote}
    \colorbox{GroupYellow}{\textbf{G4}} \textit{We can go through the features briefly. Where is the piece of paper with this terrible matrix?} [...] \\
    \colorbox{GroupBlue}{\textbf{G2}} \textit{I still don't understand which value to put. To calculate it, I need an exact value for the weighting.} \\ 
    \colorbox{GroupGreen}{\textbf{G1}} \textit{You can calculate it with this. The apprenticeship has 52\%. I believe that he [Harald] has over 25\%.} \\
    \colorbox{GroupRed}{\textbf{G3}} \textit{Yes, definitely, I mean, roughly speaking...} \\
    \colorbox{GroupGreen}{\textbf{G1}} \textit{She [Shifteh] has over 30\%. And she also has 2 minuses [...] and a plus.} \\ 
    \colorbox{GroupRed}{\textbf{G3}} \textit{That's also how I estimated it.} [...] \\
    \colorbox{GroupBlue}{\textbf{G2}} \textit{But how do you calculate it? [...] And why are there differences between the general weighting and the exact calculation? That doesn't click for me right now.}
\end{quote}

Here, the levels of detail in the explanations acted against participants' understanding by obscuring the actual feature weights, which were only accessible in level 2 of \textit{system details}. While improved navigation might solve this issue, it also shows the difficulty of simplifying information about AI systems without omitting key aspects. In the intention to provide an easier reading of the feature weights, which was an advantage in other cases, the explanations' clarity was reduced, and information was obscured. Addressing all information needs of AI novices~\cite{schmude2024information} thus leads to problems with information overload, as observed in previous work on `white-box' explanations~\cite{cheng_explaining_2019}. However, \textit{impeded understanding} alone did not mean that collaboration failed; rather, it depended on how groups dealt with these issues. Here, the key aspects were group cohesion~\cite{kerr_group_2004} and constructiveness~\cite{niculae_conversational_2016}. When the group dynamic was unfamiliar, it gave room to negative social mechanisms, such as `social loafing' (group loses motivation) and `fear of evaluation' (being criticized by others), and participants began to \textit{abandon understanding}. This suggests that when interaction between participants stopped, interaction with the explanations stopped as well. Limiting these negative dynamics and promoting positive ones must be a goal of both explanation design and setting.

\begin{figure}[h]
    \centering
    \includegraphics[width=\textwidth]{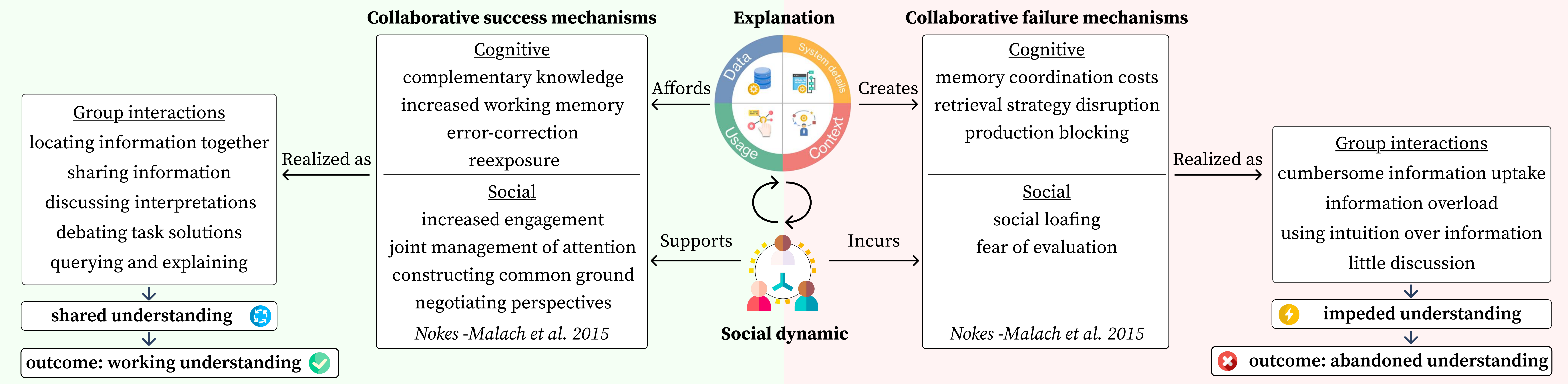}
    \caption[]{\textbf{Both explanation and social dynamic have an impact on collaborative performance.} 
    In focus groups, both explanations and social dynamic were key factors for the understanding outcome. If participants could engage easily with the explanations and each other, their interactions realized mechanisms of `collaborative success'~\cite{nokes-malach_when_2015} and led to \textit{shared understanding}. In contrast, if participants had trouble using the explanations and could not outsource or discuss these issues, interactions rather realized mechanisms of `failure'~\cite{nokes-malach_when_2015} (Section~\ref{sec:analysis}) and showed \textit{impeded understanding}. Depending on these intermediary steps, groups experienced the outcomes as \textit{working} or \textit{abandoned understanding}. From the perspective of XAI, both explanation and social dynamic are thus important aspects to keep in mind when designing explanations for groups in collaborative settings.} 
    \label{fig:findings_group_understanding}
\end{figure}

Notably, these adverse social dynamics occurred most often in Groups E and F, which were composed of job-seekers. Participants had trouble engaging with the explanations and abandoned interactions and understanding by saying: ``\textit{Probably [you can solve it] with that, but I don't know, I'm too stupid for that.}'' (E3) or ``\textit{I don't know what I should say. Everything has already been said.}'' (F2). Here, two things failed: The explanations failed to make crucial information accessible, and the group setting failed to uplift members who were discouraged. Interactions that offset this discouragement, such as \textit{locating information together} and \textit{outsourcing}, were not realized in Groups E and F. We thus propose to use co-design approaches to make explanations viable for decision subjects, as has been done with public servants~\cite{weitz2024}. Further, XAI should employ methods that create a productive social dynamic, which we identify as the second key aspect to support `collaborative success' and \textit{shared understanding} (\autoref{fig:findings_group_understanding}). 

\subsubsection{Why individuals performed better in the study tasks but still felt the absence of collaboration.}
\label{sec:results_individuals}

Despite the different advantages and disadvantages that group and individual settings offer in learning contexts (Section~\ref{sec:related_work_understanding}), these differences are rarely examined empirically in XAI\footnote{In one study that made this comparison, participants met in a chatroom and decided \textit{in collaboration with the AI} and not about it~\cite{chiang_are_2023}}. We address this gap by comparing in-person focus groups and single interviews to examine whether the social setting impacted participants' understanding. We use a triangulation\footnote{Using multiple methods of data collection (here: self-reports, task performance, articulations/actions) to explore a phenomenon~\cite{carter_use_2014}.} approach by investigating participants' understanding with respect to three aspects: interactions with the explanations, task performance, and self-reported understanding. 
% The next Section then summarizes the advantages and disadvantages of each setting in supporting AI novices' understanding. 

Overall, the explanation afforded its core functionality to both individuals and groups. Interestingly, participants in single interviews tended to request the same number or more explanations than focus groups (\autoref{fig:explanations_per_study}), despite having seemingly less working memory. S5 described: ``\textit{Well, maybe [it was overwhelming] at the very beginning [...] But I then realized that I could get through it to some extent.}'' Single interview participants also regularly \textit{consumed explanations in bulk}, i.e., read through the whole of an explanation category rapidly. This interaction was nearly nonexistent in focus groups. However, even though participants in single interviews performed better in the study tasks, they often stated that they missed the ``\textit{exchange with people, with other perspectives}'' (S3). S8 explained that this exchange would allow for a different form of understanding: 

\begin{quote}
    \textit{I think that, on your own, you can think about it very intensely and [...] make up your own mind. But that's also the disadvantage, making up your own mind. Others may have completely different thoughts and a different professional background. And that would probably have been an exciting exchange.} (S8)
\end{quote}

Participants in individual settings further performed better in the study tasks than groups (\autoref{fig:understanding_changes}). A possible explanation is that participants in single interviews engaged differently with the study tasks, as they often calculated the exact employment chance in study task 2. None of the focus groups completed this step, but rather made educated guesses. This might be explained by the degree of focused attention the settings afforded participants. As single interviews incurred no distractions, participants could immerse themselves in the explanations.
% as participant H5 commented: ``\textit{I would have liked to look through it at my own pace beforehand, [...] but then speak about it in the group afterward.}'' 

Notably, this difference is not represented in self-reported understanding. Most participants in both settings reported unchanged understanding after the explanation phase (\autoref{fig:understanding_changes}). Paradoxically, they verbally stated that it improved. E3 commented: ``\textit{I don't think I understood it the way you can understand it yet, but it's definitely better than before.}'' And S3 explained that: ``\textit{I would still say my understanding is `good', but this `good understanding' is much more informed now than the first superficial one.}'' This indicates that participants tended to judge their understanding relative to the information available, not necessarily in relation to their previous report. We describe this process as \textit{calibrating understanding}. Previous research in cognitive science has documented similar effects~\cite{keil2006}, which were also reproduced in an XAI study on white-box explanations~\cite{cheng_explaining_2019}. However, participants' verbal reports, their feedback on the explanations (Section~\ref{sec:results_explanation_design}), and the calibration process itself indicate that the explanations improved understanding. Including additional measures, such as information gain, could further capture the calibration process, which is discussed in Section~\ref{sec:discussion_design}. 

To compare individual and group interactions with the explanations, we lastly draw from the `six facets of understanding'~\cite{wiggins_understanding_2005}. The framework describes that understanding is represented by the ability to `explain, interpret, apply, take perspective, empathize, and self-reflect' with respect to a topic. The more facets are covered, the better the understanding. Seeing that individuals had a clear advantage in solving the study tasks suggests that the individual setting supported the `apply' facet. In contrast, the group settings often led participants to `explain' information to others and `interpret' it (Section~\ref{sec:results_groups_benefits}), and to `take perspective' and `empathize' through the exchange of views and experiences (Section~\ref{sec:rq2}). As explanations aim to improve understanding of a given AI system, combining both settings to cover more facets of understanding could thus be a fruitful approach. Further, explanations for individuals in particular can benefit from information covering facets usually dependent on social interaction. Our design aimed to implement this through explanations such as \textit{What chance would the system calculate for me?} (interpret -- making it personally relevant), and \textit{How could the system make the work of counselors easier or more difficult?} (take perspective: provide multiple angles and arguments). 

Based on these findings, we argue that both group and individual settings can contribute to participant understanding and should ideally be combined. In particular, focused attention can facilitate the application of information, while \textit{shared understanding} and the exchange of opinions and arguments (Section~\ref{sec:rq2}) aid encouragement, reflection, and collective action. Considering this trade-off between settings can inform how explanations can be combined with social settings to cover as many understanding facets as possible.  
%In the following Section, we reflect on the strengths and weaknesses of the explanation design and subsequently provide a summary of RQ1 in Section~\ref{sec:rq1_summary}. 

\begin{table}[h]
\caption{\textbf{Reported understanding and task performance.} This table shows participants' two understanding (und.) self-reports and performance in the four study tasks (Section~\ref{fig:study_procedure}). Increases are colored \textcolor{OliveGreen}{green}, decreases are colored \textcolor{BrickRed}{red} in reported understanding. The number of filled circles ($\bullet$) indicates the number of correct study tasks. Participants in single interviews generally performed better in the study tasks. Group H showed a particularly high task performance as participants efficiently located the relevant information together (and made an educated guess for task 2). Most other groups had a lower task performance, but still the setting acted against participant discouragement and addressed specific facets of understanding (Section~\ref{sec:rq1_summary}).}

\resizebox{\textwidth}{!}{%

% Focus groups
\begin{tabular}{cccccccccc}
\multicolumn{1}{l}{\textbf{Focus groups}} & \multicolumn{1}{l}{\textbf{}} & \multicolumn{1}{l}{\textbf{}} & \multicolumn{1}{l}{} & \multicolumn{1}{l}{} & \multicolumn{1}{l}{} & \multicolumn{1}{l}{} & \multicolumn{1}{l}{} & \multicolumn{1}{l}{} & \multicolumn{1}{l}{} \\ \hline
ID & Und. I & Und. II & Change & \multicolumn{1}{c|}{Task Performance} & ID & Und. I & Und. II & Change & Task Performance \\ \hline
\textbf{A1} & 2 & 2 & \cellcolor[HTML]{FFEB84}\textbf{0} & \multicolumn{1}{c|}{$\bullet\bullet\circ\circ$} & \textbf{F1} & 2 & 4 & \cellcolor[HTML]{98CE7F}\textbf{+2} & $\bullet\bullet\circ\circ$ \\
\textbf{A2} & 1 & 2 & \cellcolor[HTML]{CCDD82}\textbf{+1} & \multicolumn{1}{c|}{$\bullet\circ\circ\circ$} & \textbf{F2} & 1 & 4 & \cellcolor[HTML]{63BE7B}\textbf{+3} & $\bullet\bullet\circ\circ$ \\
\textbf{A3} & 4 & 4 & \cellcolor[HTML]{FFEB84}\textbf{0} & \multicolumn{1}{c|}{$\bullet\circ\circ\circ$} & \textbf{F3} & 5 & 2 & \cellcolor[HTML]{F8696B}\textbf{-3} & $\bullet\circ\circ\circ$ \\
\textbf{A4} & 4 & 4 & \cellcolor[HTML]{FFEB84}\textbf{0} & \multicolumn{1}{c|}{$\bullet\circ\circ\circ$} & \textbf{F4} & 3 & 3 & \cellcolor[HTML]{FFEB84}\textbf{0} & $\bullet\circ\circ\circ$ \\ \cline{1-5}
\textbf{B1} & 4 & 2 & \cellcolor[HTML]{FA9473}\textbf{-2} & \multicolumn{1}{c|}{$\bullet\bullet\bullet\circ$} & \textbf{F5} & 4 & 4 & \cellcolor[HTML]{FFEB84}\textbf{0} & $\bullet\bullet\circ\circ$ \\ \cline{6-10} 
\textbf{B2} & 4 & 2 & \cellcolor[HTML]{FA9473}\textbf{-2} & \multicolumn{1}{c|}{$\bullet\circ\circ\circ$} & \textbf{G1} & 5 & 5 & \cellcolor[HTML]{FFEB84}\textbf{0} & $\bullet\bullet\circ\circ$ \\
\textbf{B3} & 5 & 5 & \cellcolor[HTML]{FFEB84}\textbf{0} & \multicolumn{1}{c|}{$\bullet\bullet\bullet\circ$} & \textbf{G2} & 3 & 2 & \cellcolor[HTML]{FCBF7B}\textbf{-1} & $\bullet\bullet\bullet\circ$ \\
\textbf{B4} & 2 & 3 & \cellcolor[HTML]{CCDD82}\textbf{+1} & \multicolumn{1}{c|}{$\bullet\bullet\circ\circ$} & \textbf{G3} & 5 & 2 & \cellcolor[HTML]{F8696B}\textbf{-3} & $\bullet\bullet\circ\circ$ \\ \cline{1-5}
\textbf{C1} & 4 & 4 & \cellcolor[HTML]{FFEB84}\textbf{0} & \multicolumn{1}{c|}{$\bullet\circ\circ\circ$} & \textbf{G4} & 5 & 4 & \cellcolor[HTML]{FCBF7B}\textbf{-1} & $\bullet\bullet\circ\circ$ \\ \cline{6-10} 
\textbf{C2} & 2 & 2 & \cellcolor[HTML]{FFEB84}\textbf{0} & \multicolumn{1}{c|}{$\bullet\bullet\circ\circ$} & \textbf{H1} & 4 & 5 & \cellcolor[HTML]{CCDD82}\textbf{+1} & $\bullet\bullet\bullet\bullet$ \\
\textbf{C3} & 1 & 3 & \cellcolor[HTML]{98CE7F}\textbf{+2} & \multicolumn{1}{c|}{$\bullet\circ\circ\circ$} & \textbf{H2} & 4 & 5 & \cellcolor[HTML]{CCDD82}\textbf{+1} & $\bullet\bullet\bullet\bullet$ \\ \cline{1-5}
\textbf{D1} & 4 & 3 & \cellcolor[HTML]{FCBF7B}\textbf{-1} & \multicolumn{1}{c|}{$\bullet\bullet\circ\circ$} & \textbf{H3} & 4 & 4 & \cellcolor[HTML]{FFEB84}\textbf{0} & $\bullet\bullet\bullet\bullet$ \\
\textbf{D2} & 4 & 3 & \cellcolor[HTML]{FCBF7B}\textbf{-1} & \multicolumn{1}{c|}{$\bullet\bullet\bullet\bullet$} & \textbf{H4} & 4 & 5 & \cellcolor[HTML]{CCDD82}\textbf{+1} & $\bullet\bullet\bullet\bullet$ \\
\textbf{D3} & 4 & 4 & \cellcolor[HTML]{FFEB84}\textbf{0} & \multicolumn{1}{c|}{$\bullet\bullet\circ\circ$} & \textbf{H5} & 4 & 4 & \cellcolor[HTML]{FFEB84}\textbf{0} & $\bullet\bullet\bullet\bullet$ \\ \hline
\textbf{E1} & 3 & 4 & \cellcolor[HTML]{CCDD82}\textbf{+1} & \multicolumn{1}{c|}{$\bullet\bullet\circ\circ$} & \multicolumn{1}{l}{} & \multicolumn{1}{l}{} & \multicolumn{1}{l}{} & \multicolumn{1}{l}{} & \multicolumn{1}{l}{} \\
\textbf{E2} & 3 & 3 & \cellcolor[HTML]{FFEB84}\textbf{0} & \multicolumn{1}{c|}{$\bullet\circ\circ\circ$} & \multicolumn{1}{l}{} & \multicolumn{1}{l}{} & \multicolumn{1}{l}{} & \multicolumn{1}{l}{} & \multicolumn{1}{l}{} \\
\textbf{E3} & 4 & 4 & \cellcolor[HTML]{FFEB84}\textbf{0} & \multicolumn{1}{c|}{$\bullet\circ\circ\circ$} & \multicolumn{1}{l}{} & \multicolumn{1}{l}{} & \multicolumn{1}{l}{} & \multicolumn{1}{l}{} & \multicolumn{1}{l}{} \\ \cline{1-5}
\multicolumn{1}{l}{} & \multicolumn{1}{l}{} & \multicolumn{1}{l}{} & \multicolumn{1}{l}{} & \multicolumn{1}{l}{} & \multicolumn{1}{l}{} & \multicolumn{1}{l}{} & \multicolumn{1}{l}{} & \multicolumn{1}{l}{} & \multicolumn{1}{l}{} \\
\multicolumn{1}{l}{\textbf{Single interviews}} & \multicolumn{1}{l}{\textbf{}} & \multicolumn{1}{l}{\textbf{}} & \multicolumn{1}{l}{} & \multicolumn{1}{l}{} & \multicolumn{1}{l}{} & \multicolumn{1}{l}{} & \multicolumn{1}{l}{} & \multicolumn{1}{l}{} & \multicolumn{1}{l}{} \\ \hline
ID & Und. I & Und. II & Change & \multicolumn{1}{c|}{Task Performance} & ID & Und. I & Und. II & Change & Task Performance \\ \hline
\textbf{S1} & 5 & 5 & \cellcolor[HTML]{FFEB84}\textbf{0} & \multicolumn{1}{c|}{$\bullet\bullet\bullet\circ$} & \textbf{S7} & 4 & 4 & \cellcolor[HTML]{FFEB84}\textbf{0} & $\bullet\bullet\bullet\bullet$ \\
\textbf{S2} & 4 & 2 & \cellcolor[HTML]{F8696B}\textbf{-2} & \multicolumn{1}{c|}{$\bullet\bullet\bullet\circ$} & \textbf{S8} & 3 & 4 & \cellcolor[HTML]{CCDD82}\textbf{+1} & $\bullet\bullet\bullet\bullet$ \\
\textbf{S3} & 4 & 4 & \cellcolor[HTML]{FFEB84}\textbf{0} & \multicolumn{1}{c|}{$\bullet\bullet\bullet\bullet$} & \textbf{S9} & 5 & 5 & \cellcolor[HTML]{FFEB84}\textbf{0} & $\bullet\bullet\circ\circ$ \\
\textbf{S4} & 4 & 4 & \cellcolor[HTML]{FFEB84}\textbf{0} & \multicolumn{1}{c|}{$\bullet\bullet\circ\circ$} & \textbf{S10} & 4 & 3 & \cellcolor[HTML]{FBAA77}\textbf{-1} & $\bullet\bullet\bullet\circ$ \\
\textbf{S5} & 5 & 4 & \cellcolor[HTML]{FBAA77}\textbf{-1} & \multicolumn{1}{c|}{$\bullet\bullet\bullet\bullet$} & \textbf{S11} & 4 & 4 & \cellcolor[HTML]{FFEB84}\textbf{0} & $\bullet\bullet\bullet\bullet$ \\
\textbf{S6} & 1 & 4 & \cellcolor[HTML]{63BE7B}\textbf{+3} & \multicolumn{1}{c|}{$\bullet\bullet\bullet\circ$} & \textbf{S12} & 2 & 4 & \cellcolor[HTML]{98CE7F}\textbf{+2} & $\bullet\bullet\circ\circ$ \\ \hline
\end{tabular}}
\label{fig:understanding_changes}
\end{table}

\subsubsection{Reflections on the explanation design: Modularity, levels of detail, and most important information.}
\label{sec:results_explanation_design}

To examine how the explanation design was received, we asked participants for feedback on the explanations' structure, content, style of expression, and information coverage. We report and summarize the participants' criticisms as a basis to formulate design improvement suggestions in Section~\ref{sec:discussion}.  

\textit{Strengths and weaknesses of the design.} Positive comments described the explanations' structure as ``nicely presented'' (A2, C2) and ``good to get an overview'' (C3, H4) while being ``active and controllable'' (S8). Critical comments described the information coverage as ``too much'' (D1, S4), and the structure as ``confusing'' (B1, D1) and ``demanding'' (D2). Participants saw the design's strengths in its four-category structure, question-driven presentation, active selection, and information scope. However, the scope and depth of information also led to information overload and loss of overview. Further, the explanations' many and complicated texts were described as ``very difficult'' (E2, F2). E2 compared the language to ``\textit{letters [...] from the court. I understand every single word, but I don't understand the context.}'' Previous work has found that textual explanations can effectively convey information but tend to raise aversion with users~\cite{szymanski_visual_2021, schmude2023}. However,~\citet{weitz2024} paradoxically found that users preferred textual explanations. This points to the need for further research on textual formats in XAI, like the automated adaptation of text to different difficulties. 

\begin{figure*}[h]
    \centering
    \includegraphics[width=420pt, bb=0 0 2392 1099]{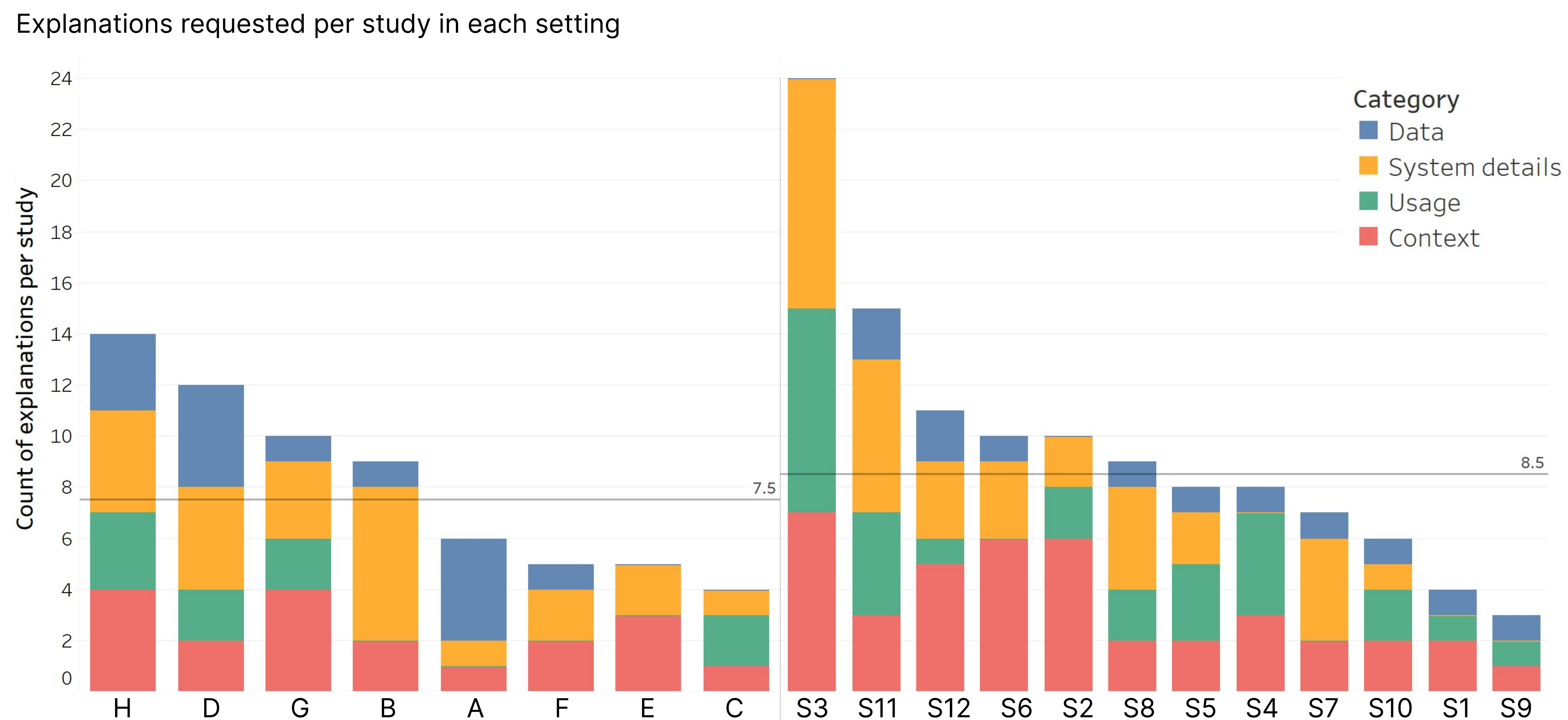}
    \caption[Explanations requested per study]{\textbf{Number of explanations requested.} The left side shows explanations requested by focus groups, the right side by participants in single interviews. The horizontal lines indicate the median. While groups were able to process many explanations by splitting the reading, several single interview participants went through equal or even higher counts. Note that \textit{context} explanations were the only category requested in every study.}
    \label{fig:explanations_per_study}
\end{figure*}

\begin{figure}[h]
    \centering
    \includegraphics[width=420pt]{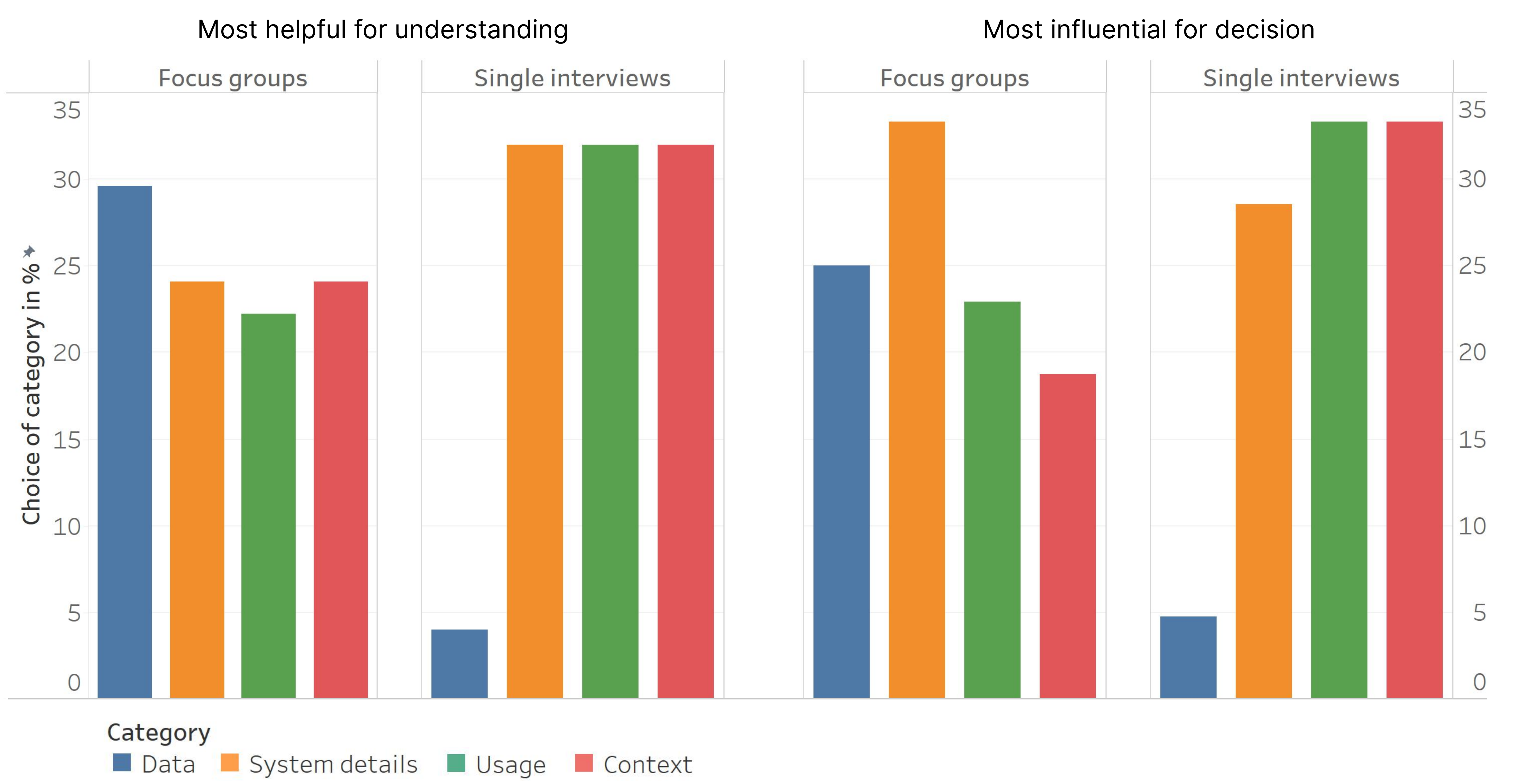}
    \caption[Most helpful explanations]{\textbf{Most helpful explanation categories for understanding and most influential categories for participants' decisions.} Participants could select any number of explanation categories for both questions, including none and all four. Focus group participants found all categories helpful for understanding, but reported \textit{system details} to be more influential for their decisions. Participants in single interviews found \textit{data} both less helpful and less influential and prioritized other categories in the given time. (Section~\ref{sec:explanation_design})} 
    \label{fig:explanations_most_helpful}
\end{figure}

\textit{Most helpful and influential information.} Participants in focus groups stated that all explanation categories helped their understanding and influenced their decision evenly (~\autoref{fig:explanations_most_helpful}), often mentioning that ``\textit{all of them [are relevant]... I don't think you can leave anything out, really}'' (D3). In contrast, participants in single interviews found \textit{data} much less helpful and less influential, stating, e.g., that they prioritized another category in the time available. Notably, participants emphasized that two categories were central: \textit{system details} and \textit{context}. \textit{System details} were perceived as ``tangible'' (S6) and ``concrete'' (S8), and explanations about the features and weighting were perceived as especially important: ``\textit{That is the central point, the basis of the whole system.}'' (G4) In turn, explanations from the category \textit{context} were requested the most (\autoref{fig:explanations_per_study}). Here, participants appreciated explanations that described decision subjects' inability to contest decisions and the system's political background. Drawing from the concept of `intelligibility types'~\cite{lim+dey_assessing_intelligibility2009}, we argue that \textit{system details} provided descriptive information to the question ``What did the system do?'', while \textit{context} provided normative information to the question ``Why did the system do [this]?''. Future research should investigate how both information types can be integrated into explanations for AI novices.

\subsubsection{\textbf{Summary RQ1}: Group and individual settings address different understanding facets in explainable AI}
\label{sec:rq1_summary}

In RQ1-Explanations, we asked how a question-driven, modular explanation design supports AI novices’ understanding in groups and individual settings. We found that the explanations support both settings but differ in how understanding develops. In groups, we found that explanations facilitated interactions that produced \textit{shared understanding} and involved cognitive and social mechanisms of `collaborative success'~\cite{nokes-malach_when_2015}. In groups with a trusting social dynamic, participants tackled explanations together, which acted against discouragement. When groups had a negative social dynamic, the explanation design could become overwhelming and understanding issues were left unchecked, leading to `process loss'~\cite {kerr_group_2004} and \textit{abandoned understanding}. Participants in single interviews interacted with the explanations in a more focused and self-directed manner. This had advantages for task performance and explanation engagement, which aligns with research on tutoring methods~\cite{bloom_2_1984}. However, the positive effects of aggregated knowledge~\cite{navajas_aggregated_2018} and peer discussion~\cite{smith_why_2009} in group settings should not be disregarded. Our findings showed that group settings can bridge understanding issues by boosting morale and letting participants share knowledge, interpretations, and experiences. We thus argue that individual and group settings support different \textit{understanding facets}~\cite{wiggins_understanding_2005}, meaning that they provide different grounds for understanding AI systems. While individual settings can make it easier to understand technical and numerical details that require much attention (`apply'), group settings can support understanding of the deployment context and consequences through the exchange of expertise and lived experiences (`interpret', `take perspective', `empathize'). Consequently, individual and group settings should be combined to leverage their different modes of interaction and understanding facets when explaining AI systems. In cases where both social settings cannot be provided, explanations of AI systems should aim to reinforce facets that are not covered in the corresponding setting. %As understanding is meant to enable action, in the following section, we examine how participants' understanding was put into action when participants deliberated the use case's deployment. 

% ======RQ2======

\subsection{RQ2-Deliberation: How do AI novices use explanations to form opinions and make decisions about AI systems in groups and individual settings?}
\label{sec:rq2}

Before and after the explanation phase, participants decided if the study's case should be deployed, and groups additionally made a collective decision (\autoref{fig:decision_table}). We compared participants' decisions and decision confidence and how they deliberated deployment to examine the impact of explanations and the social setting. 
%We found that explanations provided the basis for arguments and disagreement in group settings while also supporting single participants to engage in reasoning alone through ``internal deliberation''~\cite{mercier_reasoning_2012}. Further, we found that in one case concurrence-seeking overrode scrutiny in decision-making, in part resembling ``groupthink''~\cite{baron_so_2005, janis1971groupthink}. 
We first describe participants' confidence and decision changes in single interviews (\ref{sec:results_decisions_individuals}) and then present three cases of group discussion illustrating `elements of deliberation'~\cite{stromer-galley_measuring_2007}, including reasoned arguments (\ref{sec:results_delib_arguments}), disagreement (\ref{sec:results_delib_disagreement}), and groupthink (\ref{sec:results_delib_groupthink}). Section \ref{sec:rq2_summary} summarizes the findings.

\subsubsection{Explanation phase led to increased decision confidence and decision swings.}
\label{sec:results_decisions_individuals}

For most participants, deciding about the \amsalgorithm's deployment was a clear choice: 7 out of 8 groups and the majority of single participants voted ``No'' (\autoref{fig:decision_table}). Many participants reported increased decision confidence after the explanation phase and stated that they felt better informed due to the explanations and, where applicable, the group discussion. Reasons for these increases included a better understanding of the system's ``\textit{fundamental idea}'' (S4) and the ``\textit{exchange of different opinions and things that catch your eye}'' (G2). Participant B3 emphasized that the explanations, although they only contained factual information, provided a stark contrast to \textit{public narratives}: 

\begin{quote}
    \textit{Well, I changed my mind -- you think you understand something when you see it in the media. You have a political opinion about it. But you don't know the background information. And when you get to the background information, you can have a completely different opinion.} (B3)
\end{quote}

This contrast highlights how explanations of an AI system can impact decision-making by correcting lay understandings~\cite{devito2018} and is in line with previous work that advocates for making algorithmic complexity visible to users to tackle lay understandings~\cite{eslami2016}. It further directly connects the explanations and changes in participants' decision confidence. Few participants reported decreased decision confidence, and only S3 and S11 reported a stronger decrease of -2 (\autoref{fig:decision_table}). S3 explained that the system might have benefits, but too much depended on the \textit{conditions for deployment}. In particular, it should be deployed ``\textit{responsibly, with a pilot project, in a selected group, for three months}'', and not haphazardly, where ``\textit{you sit around for a day or eight hours and then training is finished}'' (S3). S11, who together with S3 requested the most explanations out of all participants, paradoxically stated that their confidence decreased because ``\textit{I'm still missing so much information. Especially [...] how tedious it is for the counselors if they have to disagree with the system.}'' In consequence, S11 changed their decision from `Yes' to `No'. Notably, this \textit{fear of an algorithmic imprint} was a prevalent theme throughout all studies and was often connected to past experiences with digitization projects, and the corresponding \textit{institutional deficiencies}.
%Even so, S11 changed their deployment decision from `Yes' to `No' and explained that ``\textit{it's a major fault that there's no button that says `highly motivated', which the counselor can press and then it will be counted [in the scoring]}''. 
S5, who also changed from `Yes' to `No', similarly stated that the explanations helped them to \textit{scrutinize the system}: ``\textit{You don't have to introduce anything that's extra bad}''. While the explanations thus made the \textit{decision more uncertain} for some, they undoubtedly encouraged critical reflection about the use case and triggered decision changes. Despite only having ``\textit{their own mind}'' (S8), single participants could make use of the explanations to \textit{weigh pros and cons} and \textit{adjust their mental model}. This form of ``internal deliberation'' is supported by exposure to different views, as provided through the argumentative explanations in \textit{usage} and \textit{context}. This suggests that explanations can substitute at least small parts of public deliberation, which is thought to be the more salient driver for ``reasoning towards good outcomes''~\cite{mercier_reasoning_2012}. To contrast these findings with focus groups, we illustrate this public deliberation with three conversation excerpts that showcase elements of deliberation in focus groups. 

\begin{table}[h]
\caption{\textbf{Individual and collective decisions about deploying the study's use case.} Participants were asked for their decision about the deployment of the employment prediction system before and after receiving explanations and discussing them (Section~\ref{sec:study_procedure}). Focus groups further made a collective decision about the deployment. Instances in which participants changed their votes between the first and second decision are colored \textcolor{red}{\textbf{red}}. In most focus groups, decision confidence increased after the explanation phase, with the exceptions of Groups E and F, in which participants had trouble engaging with the explanations and did not collaborate with each other (Section~\ref{sec:results_groups_drawbacks}). In single interviews, participants reported similar increases, except for S3 and S11, who explained their confidence decreases with strong adjustments to their mental models of the use case (Section~\ref{sec:results_decisions_individuals}).}

\resizebox{\textwidth}{!}{%

\begin{tabular}{cccccccccc}

% Focus groups
\multicolumn{10}{l}{\textbf{Focus   groups}} \\ \hline
ID & Decision I & Group decision & Decision II & \multicolumn{1}{c|}{Decision Conf.} & ID & Decision I & Group decision & Decision II & Decision Conf. \\ \hline
\textbf{A1} & No &  & \textcolor{red}{\textbf{Yes}} & \multicolumn{1}{c|}{\cellcolor[HTML]{FFEB84}\textbf{0}} & \textbf{F1} & Yes &  & Yes & \cellcolor[HTML]{FFEB84}\textbf{0} \\
\textbf{A2} & No &  & \textcolor{red}{\textbf{Yes}} & \multicolumn{1}{c|}{\cellcolor[HTML]{CCDD82}\textbf{+1}} & \textbf{F2} & Yes &  & Yes & \cellcolor[HTML]{FBAA77}\textbf{-1} \\
\textbf{A3} & Yes &  & Yes & \multicolumn{1}{c|}{\cellcolor[HTML]{CCDD82}\textbf{+1}} & \textbf{F3} & Yes &  & \textcolor{red}{\textbf{No}} & \cellcolor[HTML]{CCDD82}\textbf{+1} \\
\textbf{A4} & No & \multirow{-4}{*}{Yes} & \textcolor{red}{\textbf{Yes}} & \multicolumn{1}{c|}{\cellcolor[HTML]{FFEB84}\textbf{0}} & \textbf{F4} & No &  & No & \cellcolor[HTML]{FBAA77}\textbf{-1} \\ \cline{1-5}
\textbf{B1} & No &  & No & \multicolumn{1}{c|}{\cellcolor[HTML]{63BE7B}\textbf{+3}} & \textbf{F5} & Yes & \multirow{-5}{*}{No} & Yes & \cellcolor[HTML]{FBAA77}\textbf{-1} \\ \cline{6-10} 
\textbf{B2} & Yes &  & \textcolor{red}{\textbf{No}} & \multicolumn{1}{c|}{\cellcolor[HTML]{FFEB84}\textbf{0}} & \textbf{G1} & Yes &  & Yes & \cellcolor[HTML]{CCDD82}\textbf{+1} \\
\textbf{B3} & Yes &  & \textcolor{red}{\textbf{No}} & \multicolumn{1}{c|}{\cellcolor[HTML]{98CE7F}\textbf{+2}} & \textbf{G2} & No &  & No & \cellcolor[HTML]{FFEB84}\textbf{0} \\
\textbf{B4} & Yes & \multirow{-4}{*}{No} & \textcolor{red}{\textbf{No}} & \multicolumn{1}{c|}{\cellcolor[HTML]{63BE7B}\textbf{+3}} & \textbf{G3} & Yes &  & \textcolor{red}{\textbf{No}} & \cellcolor[HTML]{FFEB84}\textbf{0} \\ \cline{1-5}
\textbf{C1} & No &  & No & \multicolumn{1}{c|}{\cellcolor[HTML]{FBAA77}\textbf{-1}} & \textbf{G4} & No & \multirow{-4}{*}{No} & No & \cellcolor[HTML]{CCDD82}\textbf{+1} \\ \cline{6-10} 
\textbf{C2} & No &  & No & \multicolumn{1}{c|}{\cellcolor[HTML]{FFEB84}\textbf{0}} & \textbf{H1} & No &  & No & \cellcolor[HTML]{CCDD82}\textbf{+1} \\
\textbf{C3} & No & \multirow{-3}{*}{No} & \textcolor{red}{\textbf{Yes}} & \multicolumn{1}{c|}{\cellcolor[HTML]{98CE7F}\textbf{+2}} & \textbf{H2} & No &  & No & \cellcolor[HTML]{CCDD82}\textbf{+1} \\ \cline{1-5}
\textbf{D1} & No &  & No & \multicolumn{1}{c|}{\cellcolor[HTML]{CCDD82}\textbf{+1}} & \textbf{H3} & No &  & No & \cellcolor[HTML]{63BE7B}\textbf{+3} \\
\textbf{D2} & No &  & No & \multicolumn{1}{c|}{\cellcolor[HTML]{FFEB84}\textbf{0}} & \textbf{H4} & No &  & No & \cellcolor[HTML]{FFEB84}\textbf{0} \\
\textbf{D3} & No & \multirow{-3}{*}{No} & \textcolor{red}{\textbf{Yes}} & \multicolumn{1}{c|}{\cellcolor[HTML]{FFEB84}\textbf{0}} & \textbf{H5} & No & \multirow{-5}{*}{No} & No & \cellcolor[HTML]{CCDD82}\textbf{+1} \\ \hline
\textbf{E1} & No &  & No & \multicolumn{1}{c|}{\cellcolor[HTML]{FFEB84}\textbf{0}} &  &  &  &  &  \\
\textbf{E2} & No &  & No & \multicolumn{1}{c|}{\cellcolor[HTML]{FFEB84}\textbf{0}} &  &  &  &  &  \\
\textbf{E3} & No & \multirow{-3}{*}{No} & \textcolor{red}{\textbf{Yes}} & \multicolumn{1}{c|}{\cellcolor[HTML]{FBAA77}\textbf{-1}} &  &  &  &  &  \\ \cline{1-5}
 &  &  &  &  &  &  &  &  &  \\

% Single interviews
\multicolumn{10}{l}{\textbf{Single interviews}} \\ \hline
ID & Decision I & -- & Decision II & \multicolumn{1}{c|}{Decision Conf.} & ID & Decision I & -- & Decision II & Decision Conf.  \\ \hline
\textbf{S1} & Yes & -- & Yes & \multicolumn{1}{c|}{\cellcolor[HTML]{FFEB84}\textbf{0}} & \textbf{S7} & No & -- & No & \cellcolor[HTML]{63BE7B}\textbf{+3}  \\
\textbf{S2} & No & -- & No & \multicolumn{1}{c|}{\cellcolor[HTML]{FBAA77}\textbf{-1}} & \textbf{S8} & No & -- & No & \cellcolor[HTML]{FFEB84}\textbf{0} \\
\textbf{S3} & No & -- & No & \multicolumn{1}{c|}{\cellcolor[HTML]{F8696B}\textbf{-2}} & \textbf{S9} & No & -- & No & \cellcolor[HTML]{CCDD82}\textbf{+1}  \\
\textbf{S4} & No & -- & No & \multicolumn{1}{c|}{\cellcolor[HTML]{CCDD82}\textbf{+1}} & \textbf{S10} & Yes & -- & Yes & \cellcolor[HTML]{FFEB84}\textbf{0} \\
\textbf{S5} & Yes & -- & \textcolor{red}{\textbf{No}} & \multicolumn{1}{c|}{\cellcolor[HTML]{98CE7F}\textbf{+2}} & \textbf{S11} & Yes & -- & \textcolor{red}{\textbf{No}} & \cellcolor[HTML]{F8696B}\textbf{-2}  \\
\textbf{S6} & No & -- & No & \multicolumn{1}{c|}{\cellcolor[HTML]{98CE7F}\textbf{+2}} & \textbf{S12} & No & -- & No & \cellcolor[HTML]{CCDD82}\textbf{+1} \\ \hline

\end{tabular}}
\label{fig:decision_table}
\end{table}

\subsubsection{Case 1 - reasoned arguments: Group B discusses whether to deploy the system.}
\label{sec:results_delib_arguments}

Group B was composed of staff members and volunteers of a civil society organization. Three participants in this group changed their votes from `Yes' in the first report to `No' in the second report. We found this change to be driven by three main deliberation elements~\cite{stromer-galley_measuring_2007}: `sourcing' information, `reasoned arguments' (opinion claims grounded in the information), and `engagement' with the topic and between participants. In the excerpt, B2 and B3 \textit{weigh pros and cons} of deployment. B3 grounds their arguments in explanations about the system's features and weightings (\textit{system details}), changing the discussion's course: 

\begin{quote}
    \colorbox{GroupBlue}{\textbf{B2}} \textit{I'm skeptical, but I'm still in favor of introducing it. Because it could be an aid and a relief for the staff working there.} \\
    \colorbox{GroupRed}{\textbf{B3}} \textit{I was originally in favor for these reasons, but since I've seen these parameters, I would be very much against it. Because I think there's a lot of ideology in it. I think it's no longer acceptable that men are favored over women and that duty of care only applies to women. This comes from a time that should be long gone.} \\
    \colorbox{GroupBlue}{\textbf{B2}} \textit{Those are strong arguments.} \\
    \colorbox{GroupRed}{\textbf{B3}} \textit{The things that come out are so absurd as well. For example, Harald's apprenticeship was rated positively, but he can't even use the apprenticeship for retraining. [...] As much as I like the idea, I don't like the parameters.} \\
    \colorbox{GroupGreen}{\textbf{B1}} \textit{Did you vote yes first?} \\
    \colorbox{GroupRed}{\textbf{B3}} \textit{I ticked yes at first, but I was really shocked at what was in there [in the system].} [...] \\
    \colorbox{GroupGreen}{\textbf{B1}} \textit{What I'm wondering is, what would be the real benefit of introducing the system?} [...] \\
    \colorbox{GroupYellow}{\textbf{B4}} \textit{It's a grid, a structure for the people who work at the agency, so that they can quickly find a box.}
\end{quote}

The excerpt highlights how the explanations led B3 to \textit{change their deployment decision} and served as \textit{discussion triggers}. In the resulting discussion, participants state both arguments (\textit{discrimination}, \textit{what's the benefit?}) and opinions (\textit{disagreement with policy choices}, \textit{AI can assist in decisions}). Note that there is a difference between arguments (expression of reasoning processes that can be defended against critique) and opinions (expression of the speaker's belief)~\cite{stromer-galley_measuring_2007, mercier_reasoning_2012}. While conceiving arguments to persuade interlocutors can result in confirmation bias (interpreting evidence such that it confirms existing beliefs)~\cite{mercier_why_2011}, the fact that B3 changed their attitude, in fact, indicates that the explanations acted against this bias. We argue that the excerpt thus shows a positive synergy in that the explanations provided grounds for `arguments', which entered the discussion and provoked `collective reasoning' and three decision swings. However, considering the large argumentative influence of B3, it should also be considered how the discussion would unfold if B3 had advocated \textit{for} deployment. A case with comparable dynamics is described in~\ref{sec:results_delib_groupthink}.    

\subsubsection{Case 2 - disagreement: Group D debates normative positions regarding the algorithmic representation of people.} 
\label{sec:results_delib_disagreement}

Group D was composed of participants who had been job-seeking in the past. When the group discussed the AI system's deployment, the conversation shifted to how features that represent job-seekers' profiles should be selected and weighed. This produced disagreement, an ``important marker for deliberation''~\cite{stromer-galley_measuring_2007} that displays heterogeneity of viewpoints, acts against polarization, and involves close examination of others' reasoning. In the excerpt, D3 argues for the system's deployment, while D1 argues against, and D2 acts as a mediator:

\begin{quote}
    \colorbox{GroupRed}{\textbf{D3}} \textit{I believe that the system can form the initial basis, based on the unalterable facts, which are of course weighted, but then it has to be enriched by a human being.} [...] \\
    \colorbox{GroupGreen}{\textbf{D1}} \textit{But I don't believe that there are unalterable facts -- well, not in this area. It's all a question of representation and the lens through which you see the world.} \\
    \colorbox{GroupRed}{\textbf{D3}} \textit{When the job-seeker says, `I only have four years of elementary school', then that's four years of elementary school...} [...] \\
    \colorbox{GroupBlue}{\textbf{D2}} \textit{That doesn't mean that he can’t still be a very educated person.} \\ 
    \colorbox{GroupRed}{\textbf{D3}} \textit{But that is hard to sell to an employer, right?} [...] \\
    \colorbox{GroupBlue}{\textbf{D2}} \textit{I'm skeptical about the data. You [D3] said it's the `basis', I think there are cracks in this basis. And I'm afraid [...] that something will be pre-determined...} \\
    \colorbox{GroupRed}{\textbf{D3}} \textit{But the human decision is always subjective.} \\
    \colorbox{GroupBlue}{\textbf{D2}} \textit{That has to be weighed up. On the one hand, you have the arbitrariness of the individual employee, yes, and on the other hand, you have an incomplete picture of a person.} \\
    \colorbox{GroupGreen}{\textbf{D1}} \textit{Or a false image.} \\ 
    \colorbox{GroupBlue}{\textbf{D2}} \textit{An incomplete one, I would say.} 
\end{quote}

The group here \textit{discusses diverging views} and expresses opinions. While these opinions are meant to persuade and defend, they are grounded in \textit{lived experiences} rather than in the explanations. The central conflict develops between D1's belief that the system \textit{misrepresents reality} and D3's viewpoint that it can \textit{increase objectivity} and \textit{assist in decisions}. The discussion here did not lead to a consensus on the deployment decision in the given time, but resulted in a majority vote for `No'. We argue that it still illustrates an important process in the deliberation on public AI systems: Participants again `sourced' information that was turned into arguments, but the debate led to a more fundamental topic that surfaced discrepancies which would impede finding a collective decision. The fact that participants then engaged in `disagreement' is a sign of productive deliberation, as it shows that there were diverse viewpoints, that no polarization or `groupthink'~\cite{janis1971groupthink} occurred, and that the proposal was closely examined based on the information given~\cite{stromer-galley_measuring_2007}. In a real setting, this form of debate could serve as a fruitful basis to investigate whether the system is in the `public interest'~\cite{zuger_ai_2023} and to host `early-stage deliberations'~\cite{kawakami2024} on the system during development. The merit of this debate was further later acknowledged by D1, who found the explanations confusing but stated that these exchanges were the study's ``centerpiece'' and most intriguing part. We argue that the interplay between explanations and group discussion here supported a (simulated) evidence-informed policy-making process~\cite{mair_understanding_2019}. 

\subsubsection{Case 3 - groupthink? Group A follows a minority position and votes for system deployment}
\label{sec:results_delib_groupthink}

Group A was composed of volunteers from a civil society organization. Three participants in this group changed their decisions, shifting from `No' to `Yes' after the explanation phase. We explain these changes with three aspects: First, Group A focused on the explanation category \textit{data} and did not interact much with other categories (\autoref{fig:explanations_per_study}). This meant less attention was paid, for example, to the system's feature selection and weightings that were decisive in Cases 1 and 2. Second, participants of Group A stated that they were not directly affected by the system, as they were retired, implying low `engagement': ``\textit{It doesn't affect me anymore and I think to myself, yeah...}'' (A1). Third, participants prioritized group concurrence above a ``careful, critical scrutiny''~\cite{janis1971groupthink}. The following excerpt illustrates the tipping point for the collective decision:

\begin{quote}
    \colorbox{GroupRed}{\textbf{A3}} \textit{I still think the system is better, even if there are still mistakes in it, than sitting opposite someone [a counselor] who doesn't like you... [...] } \\
    \colorbox{GroupBlue}{\textbf{A2}} \textit{So rather `no'?}\\
    \colorbox{GroupGreen}{\textbf{A1}} \textit{Yes, as A3 says, it's... I don't know.}\\
    \colorbox{GroupRed}{\textbf{A3}} \textit{Yes and no...} [...] \\ 
    \colorbox{GroupBlue}{\textbf{A2}} \textit{I mean, it can't be avoided, it will happen. I'm convinced of that, whether we like it or not, it's done.}\\
    \colorbox{GroupGreen}{\textbf{A1}} \textit{It won't affect us anymore, at least not in the employment office. [...] I agree with the majority.}\\
    \colorbox{GroupBlue}{\textbf{A2}} \textit{But that's difficult now.}\\
    \colorbox{GroupRed}{\textbf{A3}} \textit{I'll stick with `yes'. My daughter would say I shouldn't think so negatively, especially with AI. [...]} \\
    \colorbox{GroupBlue}{\textbf{A2}} \textit{I say `yes' too. [...] You, A1 and A4, can tip the scales.}\\
    \colorbox{GroupGreen}{\textbf{A1}} \textit{I say `yes' now too, but not because I've changed my mind, but because I want an overall solution.} \\
    \colorbox{GroupYellow}{\textbf{A4}} \textit{I say `yes' but I'm leaning towards `no'.}
\end{quote}

Despite articulated reservations, all participants ultimately decided to vote for deployment. We compared this excerpt with characteristics of `groupthink', a ``mode of thinking'' in which people value concurrence higher than consideration of alternative courses of action~\cite{janis1971groupthink}. This mode produces defective decision-making processes due to three key aspects: strong social identification with the group, salient norms, and a perceived low self-efficacy to make the decision~\cite{baron_so_2005}. The excerpt clearly demonstrates two of these aspects: A1 changes their decision due to a desire for group harmony, and A4 follows suit (group identification). Further, both A1 and A3 express their uncertainty and sway between options (low self-efficacy). While A2's statement that \textit{AI is inevitable} is an opinion rather than an argument (neither `sourced' nor the product of evident reasoning), it triggers the group to make a quick decision that disregards any remaining `disagreement'. Although the process can be connected to aspects of `groupthink'~\cite{janis1971groupthink}, such as rationalizations of flawed logic and self-censorship, these aspects are not nearly as pronounced as in the literature~\cite{janis1971groupthink, Janis1972-victims_of_groupthink, baron_so_2005}. For example, the group did not share an illusion of unanimity, and the uncertainty among participants suggests no guiding salient norms. Still, as participants avoided `disagreement' and instead \textit{followed decisions of other}, the excerpt presents a suboptimal deliberation process~\cite{baron_so_2005}. In part, this can be attributed to the explanation's failure to make all fundamental information easily available and to not encourage analytical thinking over intuitive, heuristical thinking~\cite{bucinca_trust_2021}. In addition, the group might have missed a role that explicitly takes the opposing viewpoint to fuel discussion, which was identified to benefit deliberation in previous XAI research~\cite{chiang_enhancing_2024}. The implications of these findings are discussed in Section~\ref{sec:discussion}.     

\subsubsection{\textbf{Summary RQ2}}
\label{sec:rq2_summary}

The findings in this Section demonstrate how explanations supported deliberation in focus groups and single interviews. Many participants reported improved decision confidence and changed their deployment decisions based on the explanations, often due to a disillusionment regarding the \amsalgorithm's assumed merits. These changes occurred in both settings, suggesting that the explanations supported public and internal deliberation~\cite{mercier_reasoning_2012}. In group settings, participants used explanations when discussing deployment, as illustrated in Case 1. Case 2 further highlights that explanations surfaced discrepancies in personal beliefs and produced productive conflict. In contrast, Case 3 shows a deployment decision based more on concurrence-seeking than on `collaborative reasoning'~\cite{moshman_collaborative_1998}. However, we hesitate to label the exchange as `groupthink', as it does not align with all factors that characterize the phenomenon~\cite{baron_so_2005}. Based on these findings, we argue that explanations can support people in considering if AI systems are in the public interest and to discuss ``\textit{whether} and \textit{under what conditions} to move forward with developing or deploying'' them~\cite{zuger_ai_2023}. To achieve this, both the explanations and the group setting need to i) be designed so that they allow for the easy sourcing of information for arguments, ii) make all relevant information available as soon as possible, and iii) include mechanisms that encourage participants to examine both the proposal and their positions closely. Matching explanations and social setting to support `elements of deliberation'~\cite{stromer-galley_measuring_2007} thus presents promising starting points for future research on how explainable AI can promote public deliberation on AI.

\section{Discussion}
\label{sec:discussion}

% Points for discussion
% - Did understanding remain unchanged or not?
% - For which use cases would we recommend the explanation design? And in which setting should it be used?
% - What's missing from the explanation design to make it better?
% - What are the implications of all this for the real world? Where would such a combination of explanation and setting even occur?

In this section, we discuss how our findings answer our two main research questions: Whether a question-driven, modular explanation design supports AI novices' understanding in groups and individual settings (RQ1) and how AI novices used these explanations to deliberate about AI systems (RQ2). We describe the advantages of both social settings for explainable AI, outline which real-world use cases would benefit from our explanation design, discuss whether the explanations improved understanding, and provide suggestions for their design improvements. We summarize the implications of our findings in~\autoref{fig:implications}.

\subsection{Do AI novices learn and deliberate about AI better together or individually?}

In Section~\ref{sec:rq1}, we described that explanations produced \textit{shared understanding} in groups, involving both cognitive and social mechanisms of ``collaborative success''~\cite{nokes-malach_when_2015}. Section~\ref{sec:rq2} further showed that explanations improved participants' decision confidence and provided grounds for different elements of deliberation~\cite{stromer-galley_measuring_2007}, such as reasoned arguments and disagreement. In the best cases, focus groups in our study had a familiar~\cite{Johnson1985} and solution-oriented~\cite{niculae_conversational_2016} atmosphere that facilitated sharing and discussing information. In these settings, the modular explanation structure showed its strengths by allowing for the distribution of tasks among group members, providing high levels of detail and breadth if needed, and offering different viewpoints that could be used as argumentative and conversational starting points. In this sense, the explanations fulfilled their aim of supporting learning and deliberation about a public AI system~\cite{kawakami2024}. The interaction between group members is the differentiating factor compared to ``one-to-one''~\cite{naiseh_explainable_2021} explanation settings. In our study, single interviews allowed for more focused engagement with explanations and a form of ``internal deliberation''~\cite{mercier_reasoning_2012} but lacked the exchange of knowledge and perspectives with others that is deemed central for deliberation about public AI~\cite{zuger_ai_2023}. Regarding learning and deliberation, XAI would thus benefit from researching how group settings can be used to leverage collective reasoning~\cite{moshman_collaborative_1998}, wisdom of the crowds~\cite{navajas_aggregated_2018}, and performance increases through peer discussion~\cite{smith_why_2009}. However, the benefits of group settings have several preconditions, such as the containment of cognitive biases (groupthink~\cite{janis1971groupthink}, equality bias~\cite{soton2024}) and, crucially, a trusting social dynamic~\cite{chiang_are_2023}.

The importance of the social dynamic became evident in groups where members were not familiar and had trouble engaging with the explanations. In groups G and H, for example, the social dynamic bridged understanding issues of individual participants and acted against discouragement. In groups E and F, in contrast, these understanding issues eventually led participants to abandon understanding, as the social atmosphere did not support them in overcoming them. Here, a lack of trust or simply unfamiliarity between participants likely amplified effects such as social loafing and the fear of being evaluated~\cite{nokes-malach_when_2015}. This underscores the importance of creating trust between group members in collaborative XAI settings~\cite{Johnson1985}. Intuitive measures could be the introduction of a simple task that the group solves collaboratively before engaging with explanations, such as the Wason card selection task~\cite{wason1968}. Another measure could be the introduction of roles (e.g., proponents and opposition), as has been done with the ``devil's advocate'' in previous work~\cite{chiang_enhancing_2024}, to facilitate discussion and close examination of the proposal. Future work should examine how such measures can be incorporated into explanation design to support interaction in groups of comparable compositions. Lastly, regarding cognitive biases, we observed an effect resembling some aspects of ``groupthink''~\cite{janis1971groupthink} when participants in Group A changed their vote to ``Yes'' to reach a group decision. We argue that this effect originated in the lack of detailed interaction with explanations and, possibly, a perceived low degree of personal affection by the system's deployment. However, this is contrasted by participants in Group D, who debated at length about the system's deployment without reaching a consensus, despite not being directly affected. Potential measures to avoid groupthink in discussion could thus be to encourage debate, which again could be the introduction of roles to improve the ``dialectic argumentation''~\cite{mercier_why_2011}, and to explain the system in a way that makes it more personally relevant to participants~\cite{wiggins_understanding_2005}, e.g., by emphasizing connections to their own experiences.

\subsection{Did the explanations improve participants' understanding?}
\label{sec:discussion_design}

In Section~\ref{sec:rq1_summary}, we described that the explanations helped participants develop different `facets of understanding'~\cite{wiggins_understanding_2005}. In groups, participants were encouraged to `explain' information to each other and `empathize' with others' experiences, while individuals could better `apply' information in the study tasks. We further described that groups' interactions with explanations realized mechanisms of ``collaborative success''~\cite{nokes-malach_when_2015}. We thus conclude that the explanations had a positive effect on understanding. However, a more complete answer requires that we consider the difference between measurement methods and true cognitive states. In Section~\ref{sec:rq1}, we described that a majority of participants reported unchanged understanding after the explanation phase (\autoref{fig:understanding_changes}) but, paradoxically, described verbally that their understanding improved; two seemingly incongruent pieces of evidence. We explain these contradictory findings with a process we call \textit{calibrating understanding}. The term describes that participants tend to report understanding not in absolute terms, or even in relation to past understanding, but in relation to the currently available information. Participants explicitly stated that they calibrated their interpretation of 'good understanding' according to their knowledge of the information basis, which differed before and after the explanation phase. The calibration process can be traced by using concepts from the cognitive sciences: Participants i) reported their initial understanding after reading the use case introduction, they then ii) saw the explanations and realized that they had understanding gaps~\cite{rozenblit2002}, which they iii) proceeded to locate and close~\cite{keil2006}, however, they iv) also realized that they could not look at every available explanation and would develop at most a ``partial understanding''~\cite{keil_partial_understandings_2019}, which they v) rated accordingly in the second self-reports. 

Previous studies in XAI documented similar effects caused by white-box explanations~\cite{cheng_explaining_2019}, which, due to their high information density, led to increased `objective' understanding but decreased self-reported understanding. Importantly,~\cite{papenmeier_2022} found similar discrepancies when participants who received no explanation gave higher understanding scores than participants who received faithful explanations; a discrepancy our findings might explain. In the same study, similar discrepancies also occurred between trust self-reports and observations of behavior~\cite{papenmeier_2022}. In line with these findings, we argue that the calibration effect should be accounted for when measuring understanding, for example, by eliciting an additional metric that captures the perceived scope of available information. A potential reporting question could be ``How much information do you feel you currently have about the presented AI system?'', combined with a 5-point scale ranging from ``very little'' to ``very much''. Self-reported understanding could then be compared with self-reported information scope and verbal responses to acquire a more complete picture. Recent work in XAI has further proposed understanding measurement based on participants' abilities~\cite{speith_conceptualizing_2024}. This approach appears promising, as the ability to calculate study task 2 was a relevant metric in our study. We thus see eliciting understanding via multiple measures and exploring how these measures can be combined in individual and group settings as a direction for future research. 

\subsection{Which real-world settings would benefit from explainable AI in groups?}

In Section~\ref{sec:related_work_deliberation}, we described several settings where citizens gather to discuss and form opinions on matters of public interest. These included referendums, forums, and community-based spaces. This paper investigates settings suitable for deliberating the deployment of public AI systems, an issue that we frame as a matter of public interest due to the scope and severity of its potential consequences. Having established that using explanations in group settings benefits participants' understanding, decision confidence, and decision-making processes, it is worthwhile to consider how this setting could be employed in real-world contexts. One answer can be given based on participants' feedback, who stated that training in their job agency should employ a similar format to educate about AI. Notably, this feedback was given by domain experts (Groups C, G) and decision subjects (Group E), suggesting that the setting would suit both stakeholder groups for an educational intervention. Similarly, P3 explained that the explanation approach could be helpful if a similar system were used in their care facility by embedding it in the team's regular meetings, in which difficult cases are discussed and joint decisions are made. These insights are in line with previous work on XAI in public institutions. Notably,~\citet{lee_webuildai_2019} and~\citet{weitz2024} conducted participatory workshops to design explanations with end-users in the public sector, finding that co-designing explainable AI helps in considering the needs of both clients and end-users. We envision that collaborative settings and `mini-publics'~\cite{fung2003} could be useful in many contexts that aim to strengthen participatory democracy with respect to AI. Potential areas of application could be professional consultation workshops for citizens affected by algorithmic decisions, comparable to legal clinics~\cite{legalaid_ontario}, community-based education and training interventions, such as ``contestation cafés''~\cite{collins_right_2024}, or union forums that inform and organize employees' voices about the use of AI in their institution~\cite{kaur_sensible_2022}. On a different note,~\citet{crivellaro_2019} found that participatory formats that aim to connect communities to public institutions can suffer from a lack of crucial information (e.g., budgets), which could be alleviated by an information structure such as the presented explanation design. In short, we propose that explainable AI in collective settings could be a valuable engagement format for contexts in which public AI could impact people's lives. Future work could explore how collective XAI settings could be implemented in these contexts as part of responsible AI initiatives and in connection to both institutionalized~\cite{costanza-chock_who_2022} and user-based~\cite{shen_everyday_audits_2021} auditing practices.

\subsection{What's missing from the explanation design and how could it be improved?}

In Sections~\ref{sec:results_explanation_design}, we described that participants appreciated the explanations' comprehensive and flexible information selection and self-directed and active exploration. However, they also noted that the explanations have a high access threshold and require adjustment to the modular structure, making oversight difficult. A digital version of the explanation design could improve the overview through summaries and navigation while allowing for simple language options and cross-references. As the simple awareness of the scope of information also seemed to overwhelm participants, approaches to condense the scope and selection would be beneficial. An example could be a recommendation system that would suggest to participants explanations from different categories and levels of detail based on their stated interests and technical knowledge. However, we also emphasize that identifying explanation subtopics and splitting them up into levels of detail is challenging. Our explanations' structure provides different levels of soundness and completeness~\cite{kulesza_too_2013}, but deciding how the available information is allocated into this structure requires a subjective choice. In our study, detail levels 2 and 3 were supposed to convey more detailed but also more difficult explanations than the base levels. Still, some participants stated that the most critical information for them resided in level 3. We see avenues for future work in selecting and hierarchically structuring information to be included in explanations for AI systems, such that they allow for exploration while not obscuring essential information. This essential information should balance textual and visual design to ensure its uptake does not rely only on textual understanding. 

Due to the modular structure and the complexity of some of the information, several participants used intuition in answering the study tasks instead of truly searching for the explanations for solutions. Framing this in the dual-process theory of cognition, we observed that participants in these moments used System 1 (intuitive heuristics) rather than System 2 thinking (analytical reasoning)~\cite{bucinca_trust_2021, kahneman2011thinking}. \citet{lambe2016dual} listed strategies to counteract this tendency and encourage analytical thinking in the medical domain, including checklists, cognitive forcing mechanisms (consideration of alternative diagnoses, reconsideration of diagnoses), guided reflection, and use of particular reasoning approaches. \citet{bucinca_trust_2021} further tested cognitive forcing mechanisms in an AI-assisted decision-making scenario, in which they used three interventions: showing participants an AI decision only on demand, showing the AI decision only after the participants made their own predictions, and letting them wait before showing them the AI decision. These mechanisms improved the performance of the human-AI teams but led users to dislike the interface's usability, presenting a trade-off. Nonetheless, we concur that these strategies should be considered in future explanation design to ``ensure that people will exert effort to attend to those explanations''~\cite{bucinca_trust_2021}.   

\subsection{Summary of explanation design suggestions}
\label{sec:implications}

\begin{figure}[H]
    \centering
    \includegraphics[width=\textwidth, bb=0 0 905 575]{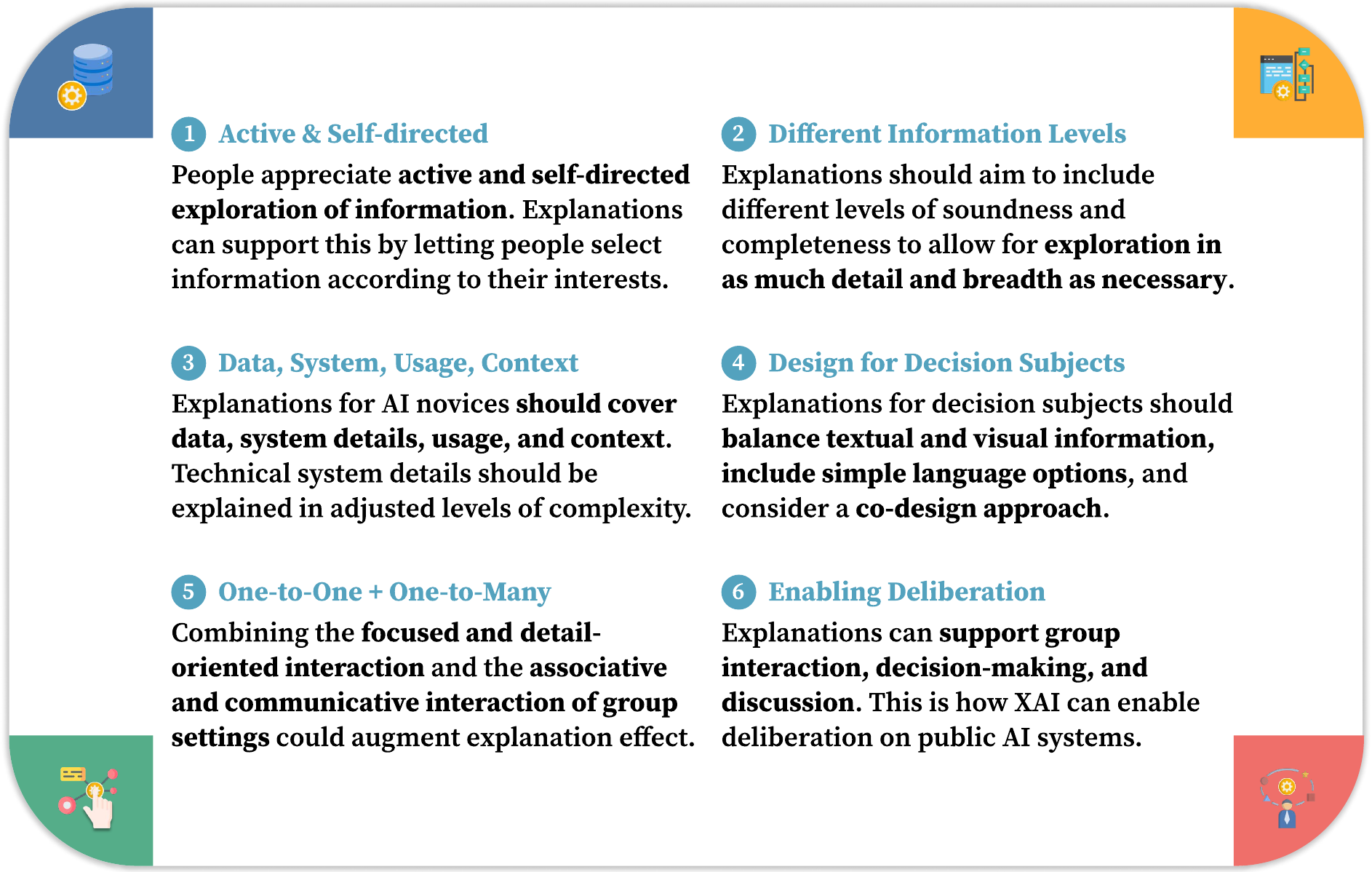}
    \caption[Implications]{\textbf{Summary of implications regarding the design of explanations for individual and collaborative settings based on our findings.}}
    \label{fig:implications}
\end{figure}

We summarize the implications of our findings in the form of suggestions for the design of explanations suited to AI novices in individual and group settings in~\autoref{fig:implications}. 

\section{Limitations}

Like any research, this study had limitations. Due to the limited sample size, we did not analyze the impact of sex and/or gender on our results, limiting the results' generalizability regarding these aspects but not their overall validity. Further, our participants were recruited from organizations and networks in the same geographical region, perhaps resulting in regional or cultural biases. The presented use case is further embedded in a specific sociotechnical context~\cite{ehsan2023} that might affect participants' understanding and perceptions (e.g., perceptions might differ between employability prediction and credit approval), and thus, a change in the domain might also change the explanations' effect. However, this does not limit the transferability of the explanation design, which can be seen as a template that can be adjusted to other use cases. We further note that our participant sample is biased toward university education in the single interviews, which we addressed by comparing these participants mostly with university-educated participants in the focus groups.  

We are further aware that the cooperation with civil society organizations in the recruitment of participants could have led to selection biases, especially in the form of convenience sampling (over-representation of readily available participants), self-selection bias (over-representation of strongly motivated participants), and interviewer bias (over-representation of agreeable or compatible participants)~\cite{collier1995translating}. We aimed to counteract these biases by defining research goals and methods clearly before recruiting participants, by using multiple recruitment sources and methods, ensuring that group composition was diversified, and by reflecting on possible sampling influences in the analysis of results. Due to the qualitative approach, our recruitment strategy further did not aim for statistical generalizability but instead intended to cover a variety of ``theoretically relevant cases''~\cite{collier1995translating} and ``careful contextualization''~\cite{collier_insights_1996} to examine our research questions. 

\section{Conclusion}

This paper tested a question-driven, modular explanation design with AI novices in groups and individual settings. We conducted an interview study involving 8 focus groups and 12 single interviews. We analyzed them to examine the effect of explanations on understanding, decisions, and decision confidence, participants' perceptions of key information, and the interaction processes in both settings. We found that explanations supported participants' understanding and decision-making differently, encouraging focused interaction in individual settings and shared understanding in group settings. Even though individuals could not exchange with others, the explanations still led to increased decision confidence and changes by supporting internal deliberation. In groups, the explanation design afforded a set of interactions that allowed participants to support each other's understanding, and further provided grounds for exchanging arguments about key aspects regarding the system's deployment. For groups that experienced collaborative failure, we suggest the modification of the explanation’s design to highlight essential information and measures to create a more productive social dynamic. With this work, we aim to showcase the potential of combining explanations with group settings to enable AI novices to understand and deliberate about public AI systems. 

% Status Quo
% What we say
% What we did
% What we showed
% What it contributes to the field
% What's next 

\begin{acks}
This work has been funded by the Vienna Science and Technology Fund (WWTF) [10.47379/ICT20058] as well as [10.47379/ICT20065]. 
\end{acks}

\bibliographystyle{ACM-Reference-Format}
\bibliography{bibliography}

\appendix
\def\thesection{\Alph{section}}

\newpage

\section*{Supplementary material \\ Better Together? The Role of Explanations in Supporting Novices in Individual
and Collective Deliberations about AI}

The supplementary material provides additional information on the study and use case. In Section~\ref{sec:employment_algo}, we provide the features used by the employment prediction algorithm, the mock newspaper article introducing the use case to participants, and the study task description. Section~\ref{sec:interview_guide} lists the self-report and interview questions that were asked during the first and second self-reports.  Section~\ref{sec:collection_explanations} depicts all explanations that participants could receive in the study. Lastly, Section~\ref{sec:code_book} provides the complete code book, listing themes and codes generated from the qualitative analysis. 

% \tableofcontents

\section{Additional information on the employment prediction algorithm}
\label{sec:employment_algo}

In this section, we give additional information on the features used by the employment prediction algorithm, the mock newspaper article introducing the system to participants, and the study task participants solved.

\subsection{Features used by the employment prediction algorithm}

\begin{table}[h]
\caption{The employment prediction algorithm uses a small set of features to calculate employability scores, including features describing demographic attributes, education, and past occupation, with "prior occupational career" being constituted by four variables. The term "cases" describes the number of times a job-seeker registered at the employment agency, "intervals" refers to a pre-defined time range, and "measures" describe support measures such as qualification courses and subsidization.}
\begin{tabular}{p{0.35\textwidth}|p{0.55\textwidth}}
\hline
Variable & $\bullet$ Nominal values \\ \hline
Gender & $\bullet$ Male/Female \\
Age group & $\bullet$ 0–29/30–49/50+ \\
Citizenship & $\bullet$ [Deployment country]/EU/Non-EU \\
Highest level of education & $\bullet$ Grade school/apprenticeship, vocational school/high- or secondary school,   university \\
Health impairment & $\bullet$ Yes/No \\
Obligations of care (only women) & $\bullet$ Yes/No \\
Occupational group & $\bullet$ Production sector/service sector \\
Regional labor market & $\bullet$ Five categories for employment prospects in assigned job center \\ \hline
Prior occupational career & $\bullet$ [Variables as described below] \\ \hline
Days of gainful employment within 4 years & $\bullet$ $<$75\%/$\geq$75\% \\
Cases within four 1-year intervals & $\bullet$ 0 cases/1 case/min. 1 case in   2 intervals/min. 1 case in 3 or 4 intervals \\
Cases with a duration longer than 180 days & $\bullet$ 0 cases/min. 1 case \\
Measures claimed & $\bullet$ 0/min. 1 supportive/min. 1 educational/min. 1 subsidized employment
\end{tabular}
\label{fig:ams_feature_table}
\end{table}

\newpage

\subsection{Mock newspaper article}

\begin{figure}[H]
    \centering
    \includegraphics[width=\textwidth]{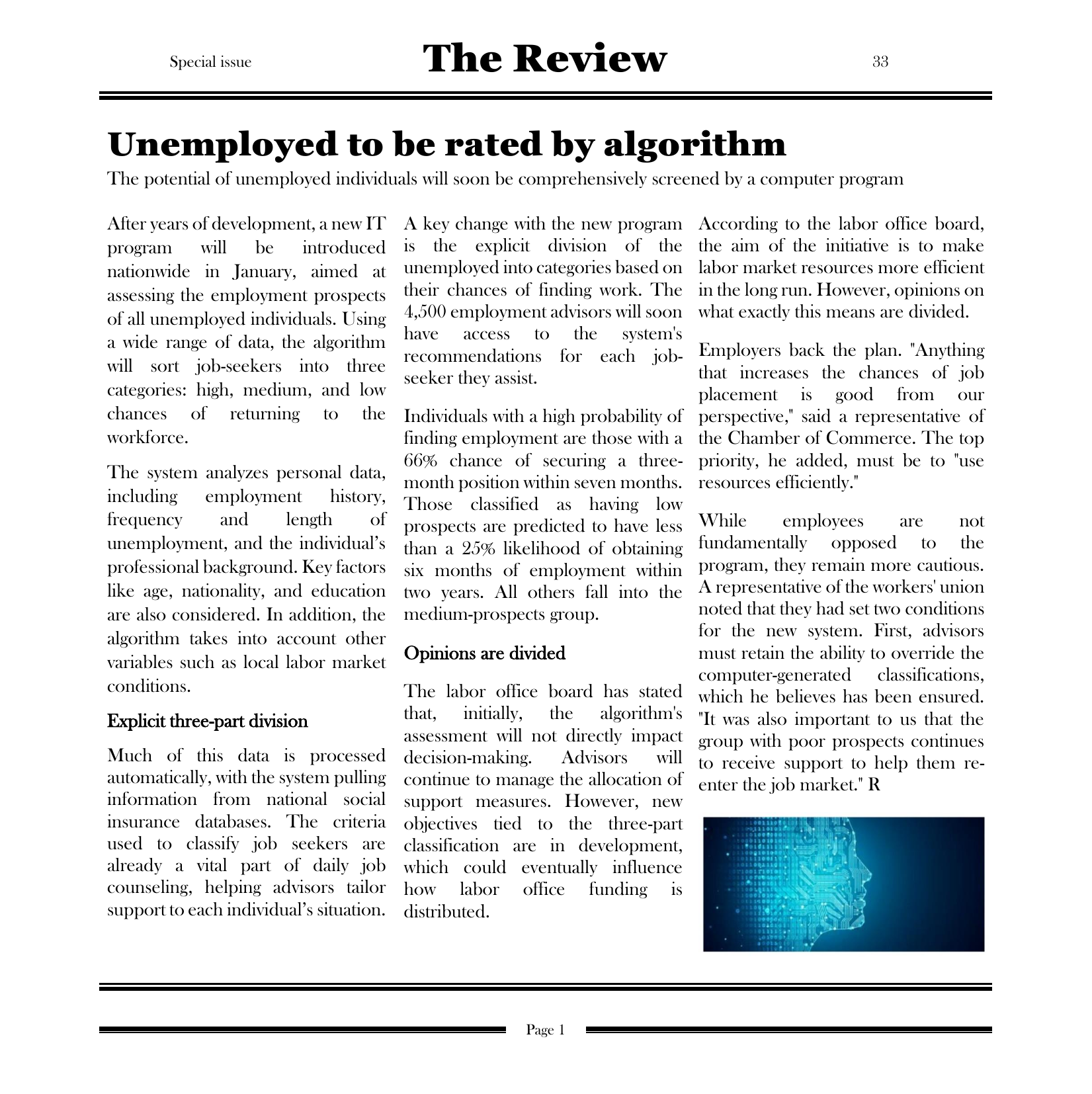}
    \caption[Mock newspaper article]{\textbf{Mock newspaper article.} Participants received initial information about the employment prediction algorithm in the form of a mock newspaper article. The article provided key information and featured the perspectives of employers and employee associations.}
    \label{fig:newspaper_article}
\end{figure}

\subsection{Task description}

\begin{figure}[H]
    \centering
    \includegraphics[width=400pt]{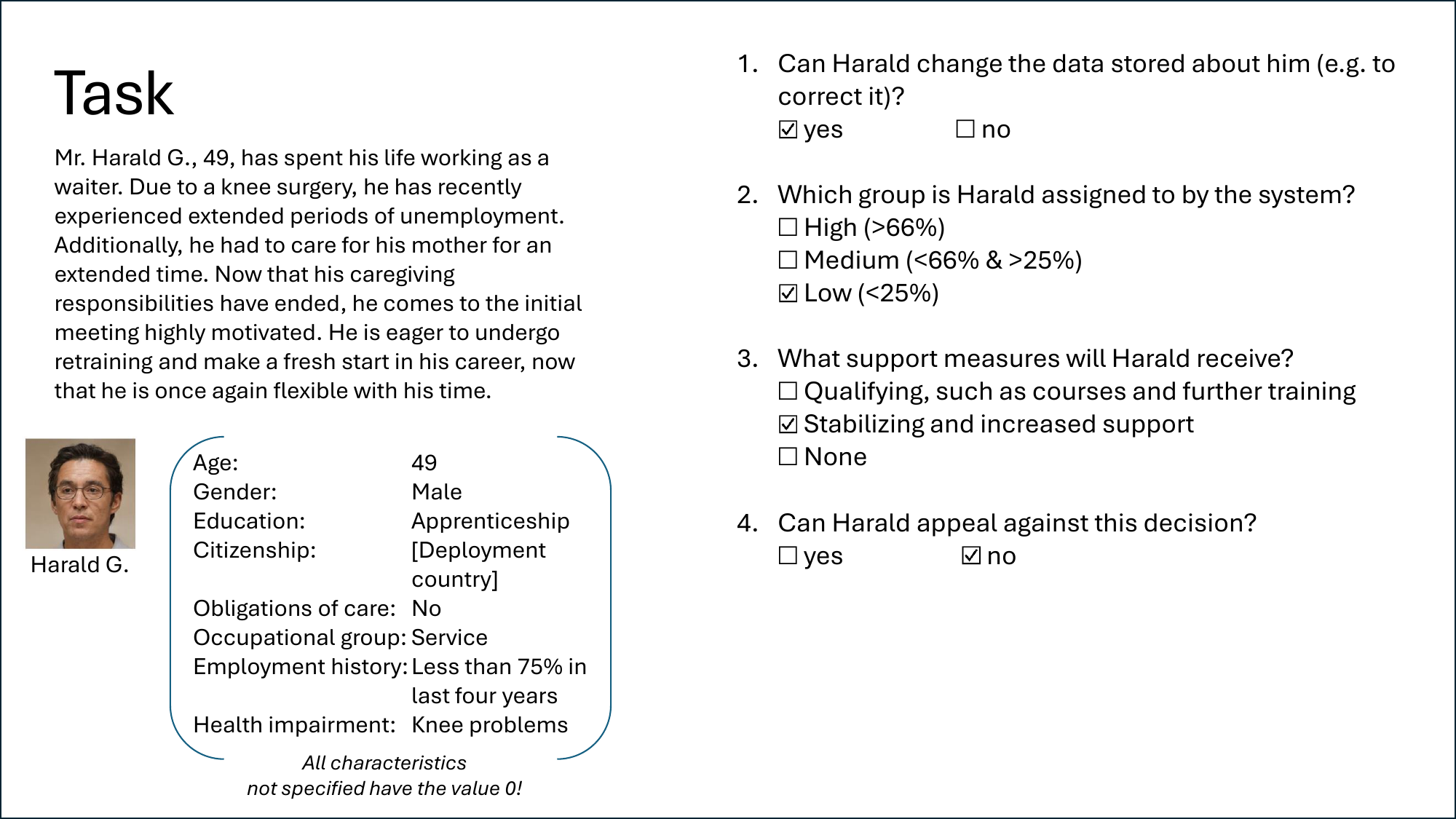}
    \caption[Study task]{\textbf{Study task.} After the first exploration phase with the explanations, participants received a fictional job-seeker case example describing Mr. Harald G.\footnote{Details on the inspiration for this case example where omitted to adhere to anonymization policy, but will be re-inserted for the final version.}: A fictional job-seeker with a brief backstory and a list of features that would be used to calculate his employment chances. Participants solved four tasks formulated as questions as depicted. The correct answers are here marked with checked boxes. Whereas tasks 1, 3, and 4 required mostly information retrieval, task 2 could be solved in two ways: by either giving an estimate based on the rough weightings in the \textit{system details} base explanations or by calculating the precise employment score. Participants had 15 (focus groups) or 20 (single interviews) minutes to solve the tasks. During that time, they could access and request all explanations and discuss possible solutions.}
    \label{fig:study_task}
\end{figure}

\newpage

\section{Self-reports and interview guide}
\label{sec:interview_guide}

\subsection{Self-reports}

Participants gave self-reports twice in the study, before and after the explanation phase (described in Section 3). In the following, we list each self-report question and the available answers.

\begin{itemize}
    \item Understanding I + II \\
    “I think that I understand the system...” \\ 
    (1 = very little; 2 = little; 3 = neither/nor; 4 = well; 5 = very well)
    
    \item Individual decision I + II \\ 
    “In your opinion, should the system be introduced?” (Yes / No)

    \item Decision confidence I + II \\    
    “In making this decision, I am...” \\
    (1 = very uncertain; 2 = uncertain; 3 = neither/nor; 4 = certain; 5 = very certain)

    \item Explanation helpfulness \\    
    "Which explanations did you find most helpful for your understanding?" \\
    (choose any from: \textit{data}, \textit{system details}, \textit{usage}, \textit{context})

    \item Explanation influence on decision \\
    "Which explanations were most influential to your decision?" \\
    (choose any from: \textit{data}, \textit{system details}, \textit{usage}, \textit{context})

    \item Contributing your voice (focus groups only) \\
    “I was able to contribute my voice in the group discussion...” \\
    (1 = very little; 2 = little; 3 = neither/nor; 4 = well; 5 = very well)

    \item Influence of discussion (focus groups only) \\
    “The group discussion influenced my decision...” \\
    (1 = very little; 2 = little; 3 = neither/nor; 4 = strongly; 5 = very strongly)

\end{itemize}

\subsection{Interview guide}

During the second self-report of participants, the investigator asked interview questions about participants' interaction with the explanations, their understanding processes, the most relevant information, and any additional situational questions. In the following, we list the questions composing the interview guide. The questions about inclusion and voice in the group were omitted in single interviews.

\begin{itemize}
    \item Understanding II
    
    \textit{How did the explanations help you to understand the system? \\ 
    What did you find difficult to understand? \\
    And how did the collaboration help you? \\
    Was something missing? An explanation or a question?}
    
    \item Individual decision II 
    
    \textit{How do you feel about this decision?}
    
    \item Decision confidence II

    \textit{How have the explanations and the collaboration influenced your decision confidence?}
    
    \item Explanation helpfulness
    
    \textit{Which of the explanations made you realize: Ah, I've understood something, that's good to know. And why?
    What effect did that have? }
    
    \textit{How did you communicate this to the group?}
    
    \item Explanation influence on discussion
    
    \textit{Which explanation made you think: Oh, that's important. It changes how I think about it. And why?}
    
    \item Contributing your voice
    
    \textit{How did you feel about the discussion process? Was everyone able to say everything?}
    
    \item Influence of discussion
    
    \textit{How do you feel about the decision the group made?}
\end{itemize}

\section{Collection of explanations}
\label{sec:collection_explanations}

\begin{figure}[H]
    \centering
    \includegraphics[width=\textwidth]{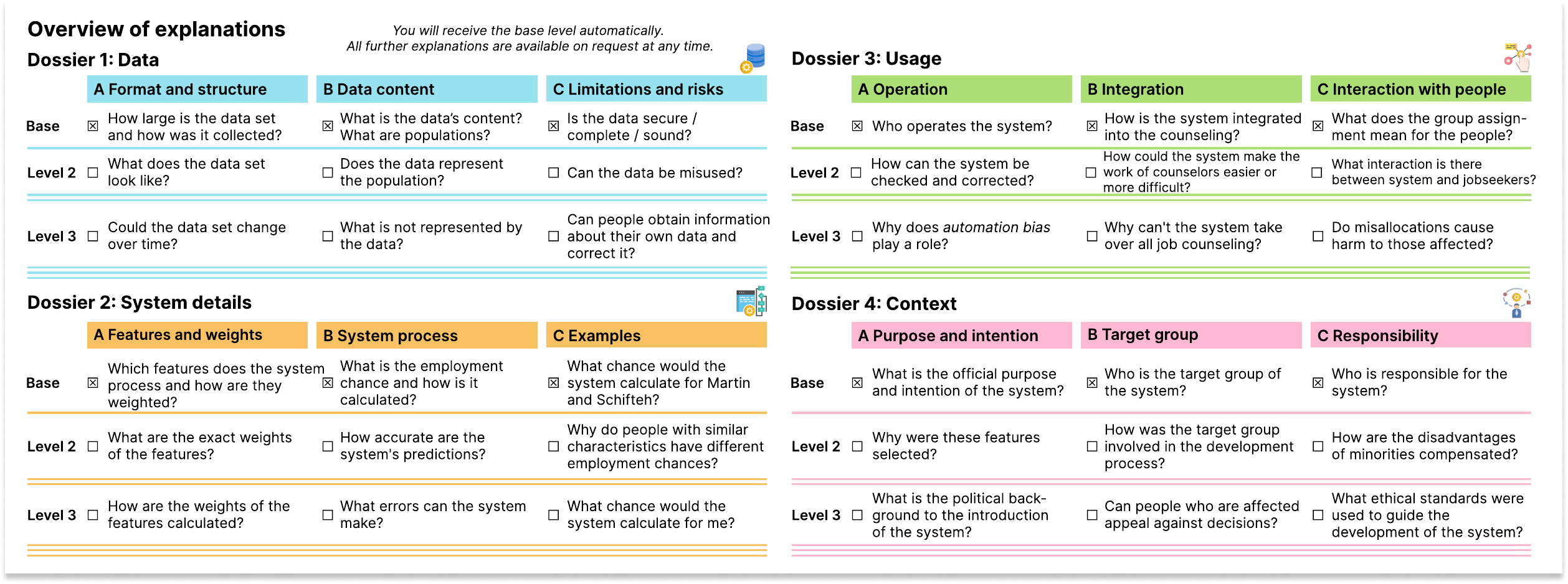}
    \caption[]{\textbf{Explanation overview.}}
    \label{fig:explanation_overview_appendix}
\end{figure}

\begin{figure}[H]
    \centering
    \includegraphics[width=300pt]{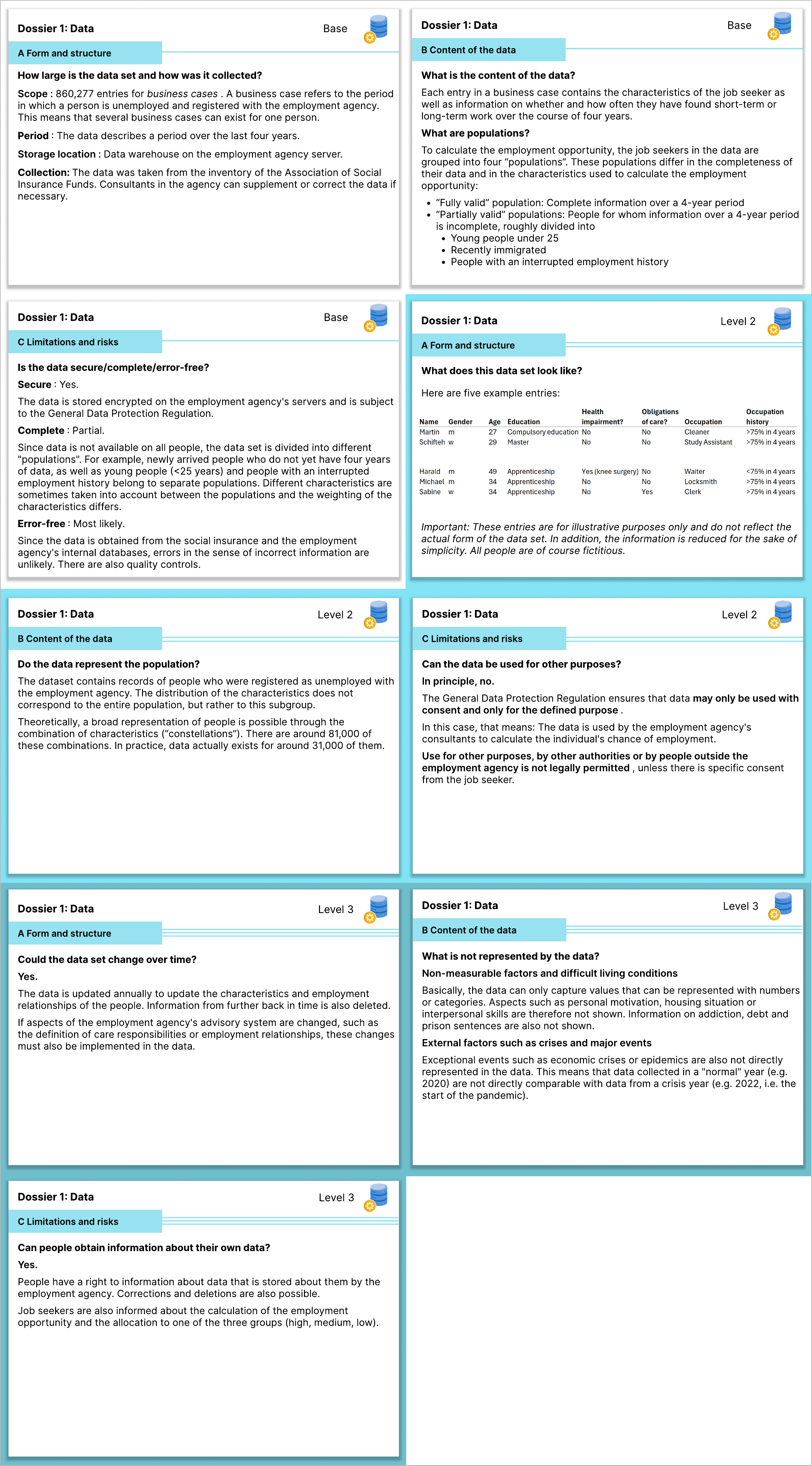}
    \caption[]{\textbf{Data.}}
    \label{fig:explanations_data}
\end{figure}

\begin{figure}[H]
    \centering
    \includegraphics[width=300pt]{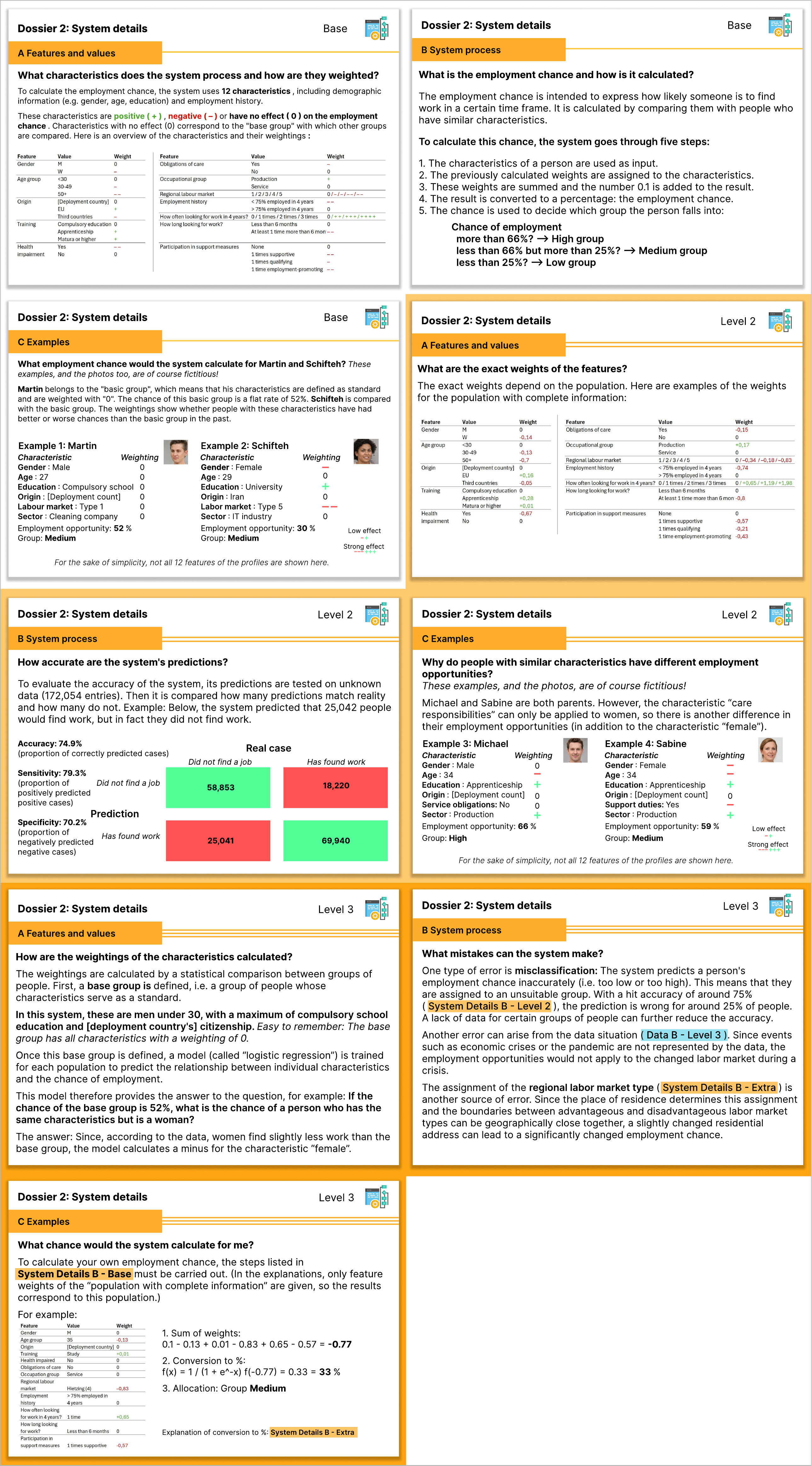}
    \caption[]{\textbf{System details.}}
    \label{fig:explanations_system_details}
\end{figure}

\begin{figure}[H]
    \centering
    \includegraphics[width=300pt]{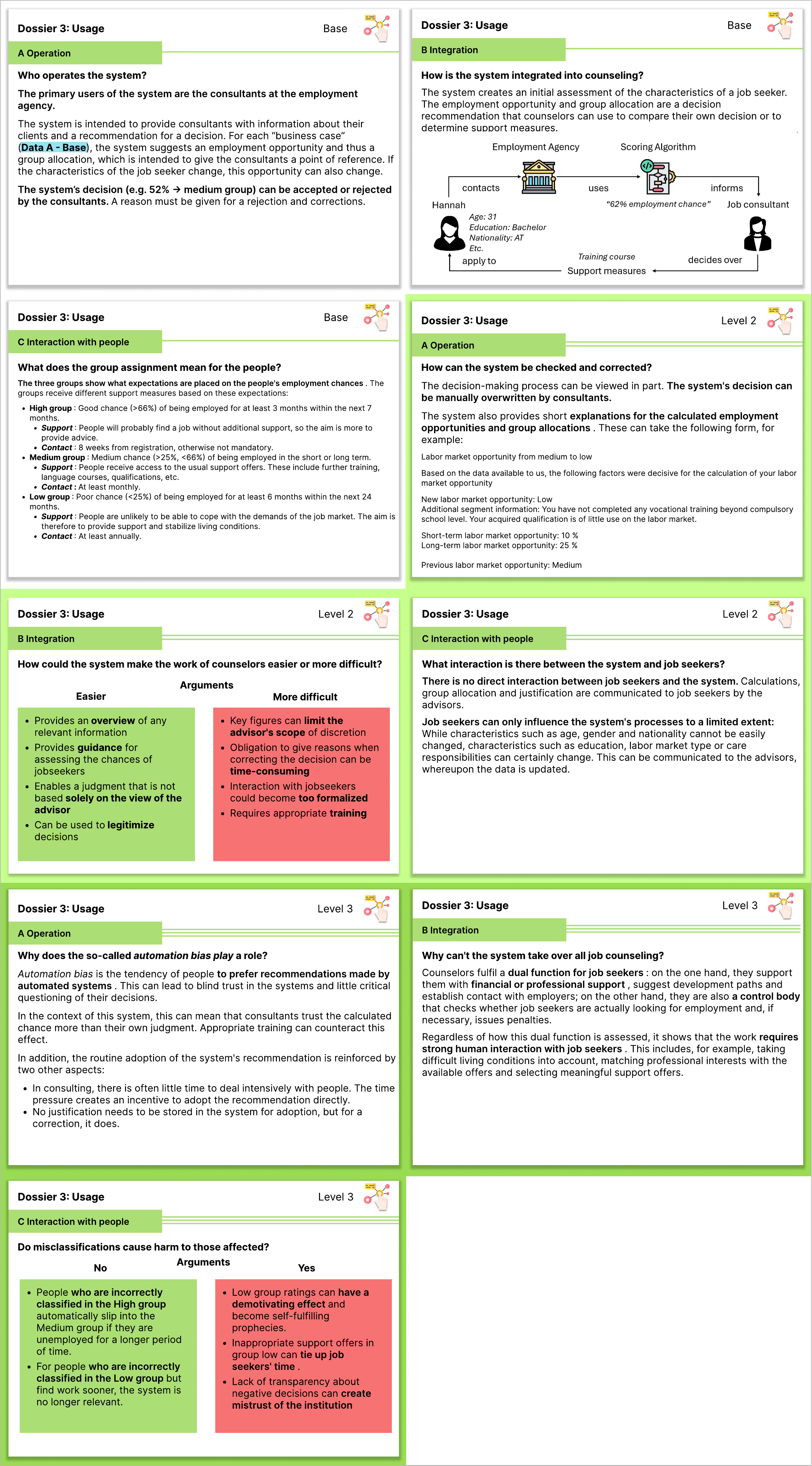}
    \caption[]{\textbf{Usage.}}
    \label{fig:explanations_usage}
\end{figure}

\begin{figure}[H]
    \centering
    \includegraphics[width=300pt]{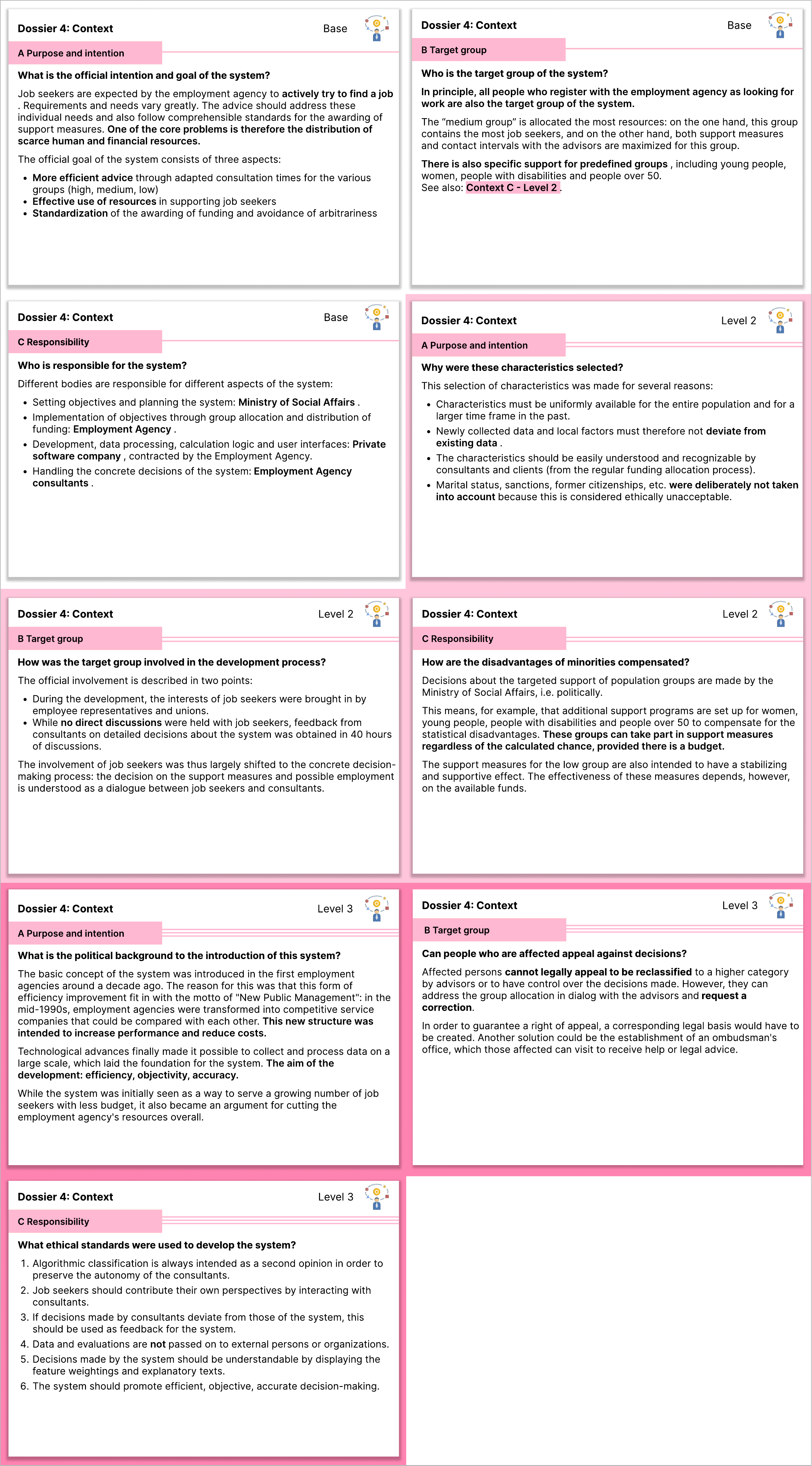}
    \caption[]{\textbf{Context.}}
    \label{fig:explanations_context}
\end{figure}

This section depicts all explanations of the four categories \textit{data} (\autoref{fig:explanations_data}), \textit{system details} (\autoref{fig:explanations_system_details}), \textit{usage} (\autoref{fig:explanations_usage}), and \textit{context} (\autoref{fig:explanations_context}). All explanation categories are split into three levels of detail; the background is colored differently for each level to facilitate distinction. Explanations of the base level were provided automatically; all others could be requested during the explanations phase. A detailed description of the explanations is given in Section 3. We insert the explanation overview again as~\autoref{fig:explanation_overview_appendix} for orientation.

\section{Code book}
\label{sec:code_book}

\autoref{fig:code_book} lists the themes and codes developed from the interview data in the qualitative analysis. The table is split into three main sections: Deliberation, understanding, and experiences and opinions. The left column lists theme groups and the right columns list single themes as coded in the data.

\begin{longtable}{l|ll}
\hline

\multicolumn{3}{l}{\textbf{Deliberation}} \\ \hline

\multicolumn{1}{l|}{Deliberation - groups} & $\bullet$ Appreciation of group   setting & $\bullet$ Forming opinions on   deployment \\
\multicolumn{1}{l|}{} & $\bullet$ Deployment decision   changed & $\bullet$ Little discussion \\
\multicolumn{1}{l|}{} & $\bullet$ Difficult to make   deployment decision & $\bullet$ No influence from group discussion \\
\multicolumn{1}{l|}{} & $\bullet$ Discussing diverging   views & $\bullet$ Strategic   decision-making \\
\multicolumn{1}{l|}{} & $\bullet$ Discussion triggers & $\bullet$ Unanimous deployment   decision \\
\multicolumn{1}{l|}{} & $\bullet$ Does not know what to   say & $\bullet$ Voting against own   interests \\
\multicolumn{1}{l|}{} & $\bullet$ Following decisions of   others & $\bullet$ Weighing pros and cons \\ \hline

\multicolumn{1}{l|}{Arguments - groups} & $\bullet$ Adverse cognitive   effects & $\bullet$ System can be misused \\
\multicolumn{1}{l|}{} & $\bullet$ Conditions for   deployment & $\bullet$ What's the benefit? \\
\multicolumn{1}{l|}{} & $\bullet$ Gaming the system & $\bullet$ What's the intention? \\
\multicolumn{1}{l|}{} & $\bullet$ Influence of Human   Factors & $\bullet$ Who can I trust? \\
\multicolumn{1}{l|}{} & $\bullet$ Integrating system into   working processes & $\bullet$ Who's in control? \\
\multicolumn{1}{l|}{} & $\bullet$ Scrutinizing the system & $\bullet$ Who's responsible? \\ \hline

\multicolumn{1}{l|}{Deliberation - individuals} & $\bullet$ AI harms trust in   institution & $\bullet$ Deployment despite   criticism \\
\multicolumn{1}{l|}{} & $\bullet$ AI will harm society & $\bullet$ Difficult to make   deployment decision \\
\multicolumn{1}{l|}{} & $\bullet$ Attitude remained   unchanged & $\bullet$ Weighing pros and cons \\
\multicolumn{1}{l|}{} & $\bullet$ Decision more uncertain   than before & $\bullet$ Would have liked a   group setting \\ \hline

\multicolumn{1}{l|}{Arguments - individuals} & $\bullet$ Conditions for   deployment & $\bullet$ Scrutinizing the system \\
\multicolumn{1}{l|}{} & $\bullet$ Decision-makers are not   the right people & $\bullet$ Supporting decision   subjects \\
\multicolumn{1}{l|}{} & $\bullet$ Is it contestable? & $\bullet$ System inherits   institutional dysfunctions \\
\multicolumn{1}{l|}{} & $\bullet$ Influence of human   factors & $\bullet$ Unions should take a   role \\
\multicolumn{1}{l|}{} & $\bullet$ Integration is the   deciding factor & $\bullet$ What are the   consequences? \\
\multicolumn{1}{l|}{} & $\bullet$ Judgment deviates from   algorithm & $\bullet$ What's the benefit? \\
\multicolumn{1}{l|}{} & $\bullet$ Lack of transparency & $\bullet$ Who's in control? \\
\multicolumn{1}{l|}{} & $\bullet$ Necessary to inform   individuals & $\bullet$ Who's responsible? \\
\multicolumn{1}{l|}{} & $\bullet$ Projecting hopes on the   AI &  \\ \hline

\multicolumn{3}{l}{\textbf{Understanding}} \\ \hline

\multicolumn{1}{l|}{Explanations - groups} & $\bullet$ Cumbersome information   uptake & $\bullet$ Incomplete coverage of   information \\
\multicolumn{1}{l|}{} & $\bullet$ Differing information   needs & $\bullet$ Interest beyond time   limit \\
\multicolumn{1}{l|}{} & $\bullet$ Exchanging explanation   sheets & $\bullet$ Locating information \\
\multicolumn{1}{l|}{} & $\bullet$ Explanation design   flaws & $\bullet$ Order of processing   information \\
\multicolumn{1}{l|}{} & $\bullet$ Explanation structure   does not make sense & $\bullet$ Overwhelmed by   information \\
\multicolumn{1}{l|}{} & $\bullet$ Explanation structure   works & $\bullet$ Piecing together   information \\
\multicolumn{1}{l|}{} & $\bullet$ Explanations help   understanding & $\bullet$ Relying on intuition   over information \\
\multicolumn{1}{l|}{} & $\bullet$ Explanations matched   with participants & $\bullet$ Requesting explanations   in bulk \\
\multicolumn{1}{l|}{} & $\bullet$ Gap between explanation   and application & $\bullet$ Suggestions for   explanation design \\ \hline

\multicolumn{1}{l|}{Understanding - groups} & $\bullet$ Abandoning   understanding & $\bullet$ Outsourcing   understanding \\
\multicolumn{1}{l|}{} & $\bullet$ Debating task solutions & $\bullet$ Participants work   individually \\
\multicolumn{1}{l|}{} & $\bullet$ Discussing   interpretations & $\bullet$ Querying and explaining \\
\multicolumn{1}{l|}{} & $\bullet$ Dividing understanding   burden & $\bullet$ Reaching working   understanding \\
\multicolumn{1}{l|}{} & $\bullet$ Impeded understanding & $\bullet$ Sharing information   with group \\
\multicolumn{1}{l|}{} & $\bullet$ Is system already in   use? & $\bullet$ Suggestions for   explanations in groups \\ \hline

\multicolumn{1}{l|}{Explanations -   individuals} & $\bullet$ Cumbersome information   uptake & $\bullet$ Explanations require   previous knowledge \\
\multicolumn{1}{l|}{} & $\bullet$ Difficult to locate   information & $\bullet$ Focus on context \\
\multicolumn{1}{l|}{} & $\bullet$ Every category is   important & $\bullet$ Focus on system details \\
\multicolumn{1}{l|}{} & $\bullet$ Explanation design   flaws & $\bullet$ Focus on usage \\
\multicolumn{1}{l|}{} & $\bullet$ Explanation design   suggestions & $\bullet$ Going into detail \\
\multicolumn{1}{l|}{} & $\bullet$ Explanations adjusted   mental model & $\bullet$ Interest beyond time   limit \\
\multicolumn{1}{l|}{} & $\bullet$ Explanations help   understanding & $\bullet$ Order of processing   information \\
\multicolumn{1}{l|}{} & $\bullet$ Explanations influenced   decision & $\bullet$ Overwhelmed by   information \\
\multicolumn{1}{l|}{} & $\bullet$ Explanations matched   participant & $\bullet$ Relying on intuition   over information \\
\multicolumn{1}{l|}{} & $\bullet$ Explanations need to   relate personally & $\bullet$ Requesting explanations   in bulk \\
\multicolumn{1}{l|}{} & $\bullet$ Explanations not suited   to decision subjects & $\bullet$ Skips category \\ \hline

\multicolumn{1}{l|}{Understanding - individuals} & $\bullet$ Calculates employment   chance & $\bullet$ Reaching a working   understanding \\
\multicolumn{1}{l|}{} & $\bullet$ Faults in algorithmic   design & $\bullet$ Understanding vs information gain \\
\multicolumn{1}{l|}{} & $\bullet$ Impeded understanding & $\bullet$ Understanding requires   example \\ \hline

\multicolumn{3}{l}{\textbf{Experiences and opinions}} \\ \hline

\multicolumn{1}{l|}{Experiences - all} & $\bullet$ AI is part of   digitization & $\bullet$ No idea of AI \\
\multicolumn{1}{l|}{} & $\bullet$ Comparing lived   experiences & $\bullet$ Overburdened   Human-in-the-Loop \\
\multicolumn{1}{l|}{} & $\bullet$ Decision subjects have   no voice & $\bullet$ Using AI at work \\
\multicolumn{1}{l|}{} & $\bullet$ Deficiencies in   institution & $\bullet$ Using AI without   knowing it \\
\multicolumn{1}{l|}{} & $\bullet$ Digital Humanism as   institutional practice & $\bullet$ Workplace wants to   integrate AI \\
\multicolumn{1}{l|}{} & $\bullet$ Is not affected by   system &  \\ \hline

\multicolumn{1}{l|}{Opinions - all} & $\bullet$ AI aversion & $\bullet$ Deploying institution   has bad reputation \\
\multicolumn{1}{l|}{} & $\bullet$ AI can assist in   decisions & $\bullet$ Disagrees with policy   choices \\
\multicolumn{1}{l|}{} & $\bullet$ AI cannot replace   humans & $\bullet$ Discrimination with and   without AI \\
\multicolumn{1}{l|}{} & $\bullet$ AI decisions must be   revisable & $\bullet$ Fears algorithmic   imprint \\
\multicolumn{1}{l|}{} & $\bullet$ AI increases   objectivity & $\bullet$ Formalization is   inevitable \\
\multicolumn{1}{l|}{} & $\bullet$ AI is inevitable & $\bullet$ Good intentions, badly   executed \\
\multicolumn{1}{l|}{} & $\bullet$ AI just appeared & $\bullet$ Need for AI-Human   collaboration \\
\multicolumn{1}{l|}{} & $\bullet$ AI misrepresents   reality & $\bullet$ No opinion \\
\multicolumn{1}{l|}{} & $\bullet$ AI openness & $\bullet$ Peaked interest \\
\multicolumn{1}{l|}{} & $\bullet$ AI replaces humans & $\bullet$ Public narratives \\
\multicolumn{1}{l|}{} & $\bullet$ AI will not improve   work &    
\label{fig:code_book}
\end{longtable}

\end{document}